\documentclass[aps,prd,twocolumn]{revtex4-2}
\usepackage{comment}
\usepackage{amsmath}
\usepackage[normalem, normalbf]{ulem}
\usepackage{color}
\usepackage{graphicx}
\usepackage{epstopdf}
\usepackage{epsfig}
\usepackage{subfig}
\usepackage{bm}
\usepackage{hyperref}
\usepackage{natbib}

\def\be {\begin{equation}}
\def\ee {\end{equation}}
\def\nn {\nonumber}
\def\bea {\begin{eqnarray}}
\def\eea {\end{eqnarray}}

\newcommand{\ep}{\epsilon}
\newcommand{\om}{\omega} 

\newcommand{\vk}{\vec k}

\newcommand{\del}{\partial}

\begin{document}
	
	\title{Effect of time-varying electromagnetic field on Wiedemann-Franz law in a hot hadronic matter}
	
	\author{Kamaljeet Singh}
	\author{Jayanta Dey}
	\author{Raghunath Sahoo\footnote{Corresponding Author: Raghunath.Sahoo@cern.ch}}
	\email{Corresponding Author: Raghunath.Sahoo@cern.ch}
	\affiliation{ Department of Physics, Indian Institute of Technology Indore, Simrol, Indore 453552, India}
	\author{Sabyasachi Ghosh}
	\affiliation{ Indian Institute of Technology Bhilai, GEC Campus, Sejbahar, Raipur-492015, Chhattisgarh, India}

	\begin{abstract}
		We have estimated the electrical and thermal conductivity of a hadron resonance gas (HRG) for a time-varying magnetic field, which is also compared with constant and
		zero magnetic field cases. Considering the exponential decay of electromagnetic fields with time, a kinetic theory framework can provide the microscopic
		expression of electrical conductivity and thermal conductivity related to baryon current in terms of relaxation and decay times. In the absence of the magnetic field, only a single time scale appears, and in the finite magnetic field case, their expressions carry two time scales - relaxation time and cyclotron time period. Estimating the conductivities
		for HRG matter in three cases - zero, constant, and time-varying magnetic fields, we have studied the validity of the Wiedemann–Franz law.
		We noticed that at a high-temperature domain, the ratio saturates at a particular value, which may be considered as Lorenz number of the hadron resonance gas.
		With respect to the saturation values, the deviation of the Wiedemann–Franz law has been quantified at the low-temperature domain. For the first time, the present work sketches this quantitative deviation of the Wiedemann–Franz law for hadron resonance gas at a constant and a time-varying magnetic field.
		
	\end{abstract}
	\date{\today}
	\maketitle
	\section{Introduction}
	Relativistic heavy ion collisions (RHICs) offer insight into the properties of deconfined matter. 
	The observation of collective behavior in the QCD plasma of the primordial particles, called quarks and gluons (QGP) has fascinated the scientific community's interest in relativistic hydrodynamics.  
	The study of dissipative hydrodynamics helps us to understand the expansion and evolution of QGP~\cite{10.1007/BFb0107310, HIRANO2006299, PhysRevC.86.014909, PhysRevC.86.014902}. The elliptic flow coefficient ($v_2$) calculated from ideal hydrodynamics is almost twice of the experimental data~\cite{PhysRevC.79.054904}. However,
	dissipative hydrodynamical calculations are in good agreement with experimental data~\cite{PhysRevLett.87.182301, PhysRevC.77.054901, PhysRevC.78.034915}. Transport coefficients like shear viscosity are essential input parameters for the simulation of dissipative hydrodynamics~\cite{PhysRevLett.99.172301, Panda:2020zhr, JAISWAL2015548}. Owing to this, microscopic calculation of transport 
	coefficients~\cite{PhysRevLett.97.152303, PhysRevC.82.024910, BOURAS2012641,PhysRevD.105.114022,Puglisi:2014pda,Plumari:2012ep,Pal:2010es,Sasaki:2008um,Ghosh:2014ija,Ghosh:2015mda,Singha:2017jmq,Abhishek:2017pkp,Ghosh:2014yea,Deb:2016myz,Marty:2013ita,Fernandez-Fraile:2009eug,Ghosh:2014qba,Ghosh:2013cba} of quark and hadronic matter became an important
	research topic in the last 10-20 years.
	According to the laws of electromagnetic theory, peripheral heavy ion collisions can generate a strong magnetic field. Recent measurements of directed flow ($v_1$) of $D^0/\bar{D}^0$ at the Large Hadron Collider (LHC) and Relativistic Heavy Ion Collider (RHIC) indicate the possibility of a massive magnetic field~\cite{PhysRevLett.125.022301, PhysRevLett.123.162301}. Refs.~\cite{MCLERRAN2014184, STEWART2021122308, PhysRevC.102.014908, PhysRevD.106.014011,Shovkovy:2022bnd,Grayson:2022asf,Tuchin:2013ie,Skokov:2009qp,Voronyuk:2011jd} and references therein
	provide good studies on the time-varying properties of this massive magnetic field, where rapidly and slowly varying both possibilities are pointed out. We considered a slowly varying magnetic field for the present work, which may or may not differ from the phenomenological picture. In that case, the hadronic phase may also
	face the magnetic field and hence, in recent times, Refs.~\cite{PhysRevD.100.114004, PhysRevD.102.016016, PhysRevC.106.044914} have found the impact of constant magnetic field on transport coefficient calculation of hadronic matter by using the hadron resonance gas (HRG) model. One can get a long list of references~\cite{PhysRevD.100.114004,HUANG20113075,PhysRevD.102.114015,PhysRevC.106.044914, PhysRevD.102.016016,Satapathy:2022xdw,Dey:2020awu,Dash:2020vxk,Nam:2012sg,Hattori:2016cnt,Hattori:2016lqx,Harutyunyan:2016rxm,Kerbikov:2014ofa,Feng:2017tsh,Wang:2020qpx,Li:2019bgc,Das:2019wjg,Das:2019ppb,Satapathy:2021wex,Li:2017tgi,Nam:2013fpa,Alford:2014doa,Tawfik:2016ihn,Tuchin:2012mf,Ghosh:2018cxb,Mohanty:2018eja,Dey:2019vkn,Hattori:2017qih,Huang:2009ue,Huang:2011dc,Agasian:2011st,Agasian:2013wta} for microscopic calculation of transport coefficients like shear viscosity, bulk viscosity, electrical, and thermal conductivity at a finite magnetic field.  The constant external magnetic field is not a realistic picture for the case of quark matter or hadronic matter produced in heavy ion collision experiments. The magnetic field can be an exponential decay function of time~\cite{STEWART2021122308,PhysRevC.102.014908,Satow:2014lia,Hongo:2013cqa}, and the present work aims to explore the transport coefficients of hadronic matter in the presence of exponentially varying magnetic field with time.
	
	Here, we have chosen two transport coefficients - electrical conductivity and thermal conductivity and their
	ratio to check the validity of the Wiedemann–Franz (WF) law.
	Electrical conductivity is one of the essential properties in determining the electromagnetic response of a medium. Lower the conductivity, the faster the decay of the electromagnetic field. Because QGP is an expanding medium, its properties, including electrical conductivity, change as it evolves. As a result, it is very complicated to formulate~\cite{STEWART2021122308,PhysRevC.102.014908} the exact decay function of the electromagnetic field in the medium,
	because of the dynamically evolving system with highly complex physical processes involving different degrees of freedom. Many studies have been done to find the decay profile of the fields. 
	Thermal conductivity is another important transport coefficient that determines how fast thermal equilibrium can be achieved in the medium. QGP created at RHIC energy can have non-zero baryon chemical potential \cite{KrzysztofRedlich_2004}. In such a system, thermal conductivity contributes to total dissipation~\cite{GAVIN1985826, PhysRevD.94.094002}. In the presence of a magnetic field, thermal conductivity also gets affected because of the presence of electrically charged particles in the medium, which are quarks in the deconfined QGP phase and charged hadrons in the confined hadronic phase~\cite{PhysRevD.100.114004, PhysRevD.102.016016, PhysRevC.106.044914}.  
	
	In this work, we estimate the thermal and electrical conductivity of hadronic matter in a time-varying electromagnetic field at finite baryon chemical potential. These transport properties in a time-varying electromagnetic field have recently been studied in Refs.~\cite{PhysRevD.106.034008, PhysRevD.104.094037} for the QGP phase, using a quasiparticle-based model. Very recently, Ref.~\cite{Ghosh:2022vjp} has studied the electrical properties of hadronic matter under a time-varying electric and constant magnetic field using the linear sigma model (LSM). We use the ideal hadron resonance gas (HRG) model for quantitative estimations in the current study. Moreover, we investigated the validity of WF law in the presence of a constant and time-varying electromagnetic field for the HRG matter. 
	
	The paper is organized in the following manner. Introducing the origin of the electromagnetic field and its effect on the transport properties of the medium, we discuss the derivation of electrical and thermal conductivity in the formalism section (\ref{sec:Formalism}). We also briefly discuss the calculation of relaxation time in the HRG medium. In section (\ref{sec-result}), we have discussed the results in detail. The thermal and electric transport properties of HRG matter are studied in the context of WF law. In section (\ref{sec-summary}), we have summarized the study with a possible outlook. Detailed calculation of the conductivities concerning the formalism section (\ref{sec:Formalism}) is given in the appendices.

	\section{Formalism}
	\label{sec:Formalism}
	In this section, we calculate the thermal and electrical conductivities of a relativistic fluid in the presence of an external time-dependent electric and magnetic field.
	The appendices contain the detailed calculation. For more detail, the reader can go through Ref.~\cite{PhysRevD.104.094037} for electrical conductivity and Ref.~\cite{PhysRevD.106.034008} for thermal conductivity. In strong field case, one should consider the Landau quantization, which is
		ignored in present work. So present work may be considered as weak field approximation estimations.
	
	\subsection{Electrical conductivity}
	\label{subsec:EC}
	In the presence of an electromagnetic field, the general form of electric current density can be expressed as
	\begin{align}\label{23}
	\vec{j} = j_e \hat{{e}} +j_H (\hat{e} \times \hat{{b}})~, 
	\end{align}
	where $\hat{e}$, $\hat{b}$ are unit vectors along the direction of electric field $(\vec{E}=E\hat e)$ and magnetic field $(\vec{B}=B \hat b)$, respectively. $j_e$ represents the  Ohmic current density along the direction of the electric field, and  $j_H$ is the Hall current density perpendicular to the electric and magnetic fields. Now, we consider a system of relativistic fluid consisting of particles with energy $\om_i = \sqrt{\vec k_i^2+m_i^2}$, momentum $\vec{k_i}$, mass $m_i$, and chemical potential $\mu_i = {\rm b}_i \mu_B$ for $\textit{i}$th species, with b$_i$ as baryon quantum number and $\mu_B$ as baryon chemical potential.
	Here species stand for different hadrons in the hadron resonance gas (HRG) model. 
	Single particle distribution function at equilibrium for $\textit{i}$th species is
	\begin{align}\label{Dis-f}
	f^0_i= \frac{1}{e^{\frac{\om_i-{\rm b}_i\mu_B}{T}}\pm 1}~,
	\end{align}
	where $\pm$ stands for fermion and boson, respectively. The total single-particle distribution function ($f_i$) for a system slightly out of equilibrium ($\delta f$) can be written as $f_i=f^0_i+\delta f_i$. In kinetic theory, electric current density for such a system can be expressed as
	\begin{align}\label{Cur-den}
	\vec{j} =  \sum_{i} q_i g_i \int \frac{d^3|\vk_i|}{(2\pi)^3} \frac{\vec{k_i}}{\om_i} f_i~. 
	\end{align}
	Here, $q_i$ is the electric charge, and $g_i$ is the degeneracy of the $\textit{i}$th species particles. Net non-zero currents arise when the system is out of equilibrium. To find the expression of $\delta f_i$, we solve the Boltzmann transport equation (BTE) with the help of relaxation time approximation (RTA). In the presence of an external electromagnetic field,  BTE under RTA can be expressed as\cite{PhysRevD.102.016016}
	\begin{align}\label{BTE-RTA}
	\frac{\del f_i}{\del t} + \frac{\vk_i}{\om_i} \cdot \frac{\del f_i}{\del \vec{x}} + q_i \left(\vec{E} + \frac{\vk_i}{\om_i} \times \vec{B}\right) \cdot \frac{\del f_i}{\del \vec{k_i}}
	= -\frac{\delta f_i}{\tau^i_R}~,
	\end{align}
	where $\tau^i_R$ is the relaxation time of the particle. To solve the above equation, we have to assume an ansatz of $\delta f_i$. The deviation of the distribution function from equilibrium is driven by the electromagnetic field, so leading order contribution in $\delta f_i$ can be assumed as~\cite{PhysRevD.104.094037}
	\begin{align}\label{delta-f}
	\delta f_i = (\vec{k_i} \cdot \vec{\Omega}_\sigma) \frac{\del f^0_i}{\del \om_{i}}~.
	\end{align}  
	The unknown vector ${\vec \Omega}_\sigma$ must be derived by the electromagnetic field, and a general form can be assumed as (up to first-order in time derivative)
	\begin{align}\label{Omega}
	\vec{\Omega}_\sigma = &~\alpha_1 \vec{E} + \alpha_2 \dot{\vec E} +\alpha_3 \vec{B} + \alpha_4 \dot{\vec B} + \alpha_5 (\vec{E} \times \vec{B}) \nn \\
	& + \alpha_6 (\dot{\vec E} \times \vec{B}) + \alpha_7 (\vec E \times \dot{\vec{B}})~. 
	\end{align}  
	In the case of second-order magnetohydrodynamics, more
		coefficients arise from the second-order derivative of fields~\cite{PhysRevD.99.056017},
		which is not considered in the present study. 
	Here, $\alpha_j$ ($j=1,2,..7$) are unknown coefficients that determine the strength of the respective field in driving the system out of equilibrium. 
	Here, we consider the case where the chiral chemical potential is zero. So, the terms $\vec{B}$, $\dot{\vec B}$ do not contribute to the current~\cite{Satow:2014lia}. Therefore, we get five components of the current density corresponding to $\alpha_1, ~\alpha_2, ~\alpha_5, ~\alpha_6, ~\alpha_7$. Whereas, the coefficients $\alpha_1, ~\alpha_2$ contribute to the Ohmic current and $\alpha_5, ~\alpha_6, ~\alpha_7$ contribute to the Hall current.
	
	Now, we consider a time-dependent electric and magnetic field of the from~\cite{Satow:2014lia, Hongo:2013cqa}
	\begin{align}
	B = B_0 \exp{\left(-\frac{t}{\tau_B}\right)},\label{Mag-Field}\\
	E = E_0 \exp{\left(-\frac{t}{\tau_E}\right)},\label{Mag-Field1}
	\end{align}
	where $B_0, E_0$ are the magnitudes of the initial fields having decay parameters of $\tau_B$ and $\tau_E$, respectively, and $t$ is the proper time. The initial value of the fields can be obtained from the impact parameter and size of the colliding nuclei~\cite{Skokov:2009qp}. For a typical gold-gold collision with an impact parameter of $\sim 7~$fm, the initial magnetic field will be of the order of $\sim 10$m$_\pi^2$~\cite{Satow:2014lia}. The decay parameters depend on the medium's properties, such as electrical conductivity, which can be fixed from magneto-hydrodynamic simulation.
	For a slowly varying magnetic field, we can consider that the inverse of cyclotron frequency is approximately equal to the magnetic field decay parameter, i.e.,  $\tau_B \approx \frac{\om_i}{q_i B}$. 
	Therefore, in the weak field limit, we can solve Eq.~(\ref{BTE-RTA})  using Eqs.~(\ref{delta-f}) and (\ref{Omega}), and get the expression of $\alpha_j$'s. The components of current density corresponding to $\alpha_1, ~\alpha_2, ~\alpha_5, ~\alpha_6$, and $ \alpha_7$ can be obtained from Eq.~(\ref{Cur-den}) as (see Appendix~\ref{appendix1})  
	\begin{align}\label{Elec-Cur}
	\nn &j_e^{(0)} = \frac{E}{3T} \sum_i g_i (q_{i})^2 \int \frac{d^3|\vk_i|}{(2\pi)^3}
	\frac{ \vk^2_i}{\om_i^2} f^0_i(1\mp f^0_i) ~ \tau_R^i \\
	\nn &~~~~~~~~~~~~~~~~~~~~~~~~~~~~ \times \frac{1}{1+\chi_i+\chi_i^2}, \\ 
	\nn &j_e^{(1)} = -\frac{\dot{E}}{3T} \sum_i g_i (q_{i})^2 \int \frac{d^3|\vk_i|}{(2\pi)^3}
	\frac{ \vk^2_i}{\om_i^2} f^0_i(1\mp f^0_i)~ {\tau_R^i}^2\\
	\nn &~~~~~~~~~~~~~~~~~~~~~~~~~ \times \frac{1+\chi_i-\chi_i^2}{(1+\chi_i)(1+\chi_i^2)(1+\chi_i+\chi_i^2)}, \\
	\nn &j_H^{(0)} = \frac{E B}{3T} \sum_i g_i (q_{i})^3 \int \frac{d^3|\vk_i|}{(2\pi)^3}
	\frac{ \vk^2_i}{\om_i^3} f^0_i(1\mp f^0_i)~ {\tau_R^i}^2\\
	\nn &~~~~~~~~~~~~~~~~~~~~~~~~~~~~ \times \frac{1}{(1+\chi_i)(1+\chi_i+\chi_i^2)}, \\ 
	\nn &j_H^{(1)} = -\frac{\dot{E}B}{3T} \sum_i g_i (q_{i})^3 \int \frac{d^3|\vk_i|}{(2\pi)^3}
	\frac{ \vk^2_i}{\om_i^3} f^0_i(1\mp f^0_i) ~ {\tau_R^i}^3\\
	\nn &~~~~~~~~~~~~~~~~~~~~~~~~~ \times \frac{\chi_i}{(1+\chi_i)(1+\chi_i^2)(1+\chi_i+\chi_i^2)}, \\
	&j_H^{(2)} = -\frac{E\dot{B}}{3T} \sum_i g_i (q_{i})^3 \int \frac{d^3|\vk_i|}{(2\pi)^3}
	\frac{ \vk^2_i}{\om_i^3} f^0_i(1\mp f^0_i) ~ {\tau_R^i}^3 \nn\\
	&~~~~~~~~~~~~~~~~~~~~~~~~~~~~ \times \frac{1}{(1+\chi_i)(1+\chi_i+\chi_i^2)}~,
	\end{align}
	where $\chi_{i} = \frac{\tau_R^i}{\tau_B}$.
	Here, the Ohmic current density is $j_e=j_e^{(0)}+j_e^{(1)}$ and Hall current density is $j_H=j_H^{(0)}+j_H^{(1)}+j_H^{(2)}$.
	
	Now, from Ohm's law, we can write
	\begin{align}\label{Ohm}
	j^{i} = \sigma^{ij} E_j~,
	\end{align}
	where $\sigma^{ij}$ is the conductivity tensor. 
	In general, it contains contributions from all the first and higher-order derivatives of the fields. Now, from Eqs.~\ref{23} and (\ref{Ohm}), we can write
		\begin{equation}
		\vec {j} = \sigma_e \vec E + \sigma_H (\vec E \times \hat{b}),
		\end{equation}
		where Ohmic ($\sigma_e$) and Hall ($\sigma_H$) conductivity components are
	\begin{align}\label{C-EC}
	&	\sigma_e=\frac{j_e}{E} = \frac{j^{(0)}_e+j^{(1)}_e}{E}
	=\sigma_e^{(0)}+\sigma_e^{(1)},\nn \\
	&	\sigma_H=\frac{j_H}{E} = \frac{j^{(0)}_H+j^{(1)}_H+j^{(2)}_H}{E}
	= \sigma_H^{(0)}+\sigma_H^{(1)}+\sigma_H^{(2)}.
	\end{align}
	The conductivity coefficients defined above are possible for the specific form of electric and magnetic field given in Eqs.~(\ref{Mag-Field}) and (\ref{Mag-Field1}), respectively.
	From Eqs.~(\ref{C-EC}) and (\ref{Elec-Cur}), components of electrical conductivity can be found as (in the weak field limit)
	\begin{align}
	\sigma^{(0)}_e &= \frac{1}{3T} \sum_i g_i (q_{i})^2 \int \frac{d^3|\vk_i|}{(2\pi)^3}
	\frac{ \vk^2_i}{\om_i^2} f^0_i(1\mp f^0_i) ~ \tau_R^i \nn\\
	\nn &~~~~~~~~~~~~~~~~~~~~~~~~~~~~ \times \frac{1}{1+\chi_i+\chi_i^2}, \\
	\sigma^{(1)}_e &= \frac{1}{3T} \sum_i g_i (q_{i})^2 \int \frac{d^3|\vk_i|}{(2\pi)^3}
	\frac{ \vk^2_i}{\om_i^2} f^0_i(1\mp f^0_i) ~\frac{{\tau_R^i}^2}{\tau_E} \nn\\
	\nn &~~~~~~~~~~~~~~~~~~~~~~~ \times \frac{1+\chi_i-\chi_i^2}{(1+\chi_i)(1+\chi_i^2)(1+\chi_i+\chi_i^2)}, \\
	\sigma^{(0)}_H &= \frac{1}{3T} \sum_{i} g_i (q_{i})^2 \int \frac{d^3|\vk_i|}{(2\pi)^3}
	\frac{ \vk^2_i}{\om_i^2} f^0_i(1\mp f^0_i) ~{{\tau_R^i}^2 \Gamma_i} \nn\\
	\nn &~~~~~~~~~~~~~~~~~~~~~~~~~~~~ \times \frac{1}{(1+\chi_i)(1+\chi_i+\chi_i^2)}, \\
	\sigma^{(1)}_H &= \frac{1}{3T} \sum_{i} g_i (q_{i})^2 \int \frac{d^3|\vk_i|}{(2\pi)^3}
	\frac{ \vk^2_i}{\om_i^2} f^0_i(1\mp f^0_i) \frac{{\tau_R^i}^3 \Gamma_i}{\tau_E} \nn\\
	\nn &~~~~~~~~~~~~~~~~~~~~~~~~~ \times \frac{\chi_i}{(1+\chi_i)(1+\chi_i^2)(1+\chi_i+\chi_i^2)},\\ 
	\sigma^{(2)}_H &= \frac{1}{3T} \sum_{i} g_i (q_{i})^2 \int \frac{d^3|\vk_i|}{(2\pi)^3}
	\frac{ \vk^2_i}{\om_i^2} f^0_i(1\mp f^0_i) \frac{{\tau_R^i}^3 \Gamma_i}{\tau_B} \nn\\
	&~~~~~~~~~~~~~~~~~~~~~~~~~ \times \frac{1}{(1+\chi_i)(1+\chi_i+\chi_i^2)}~,
	\label{ele-cond}
	\end{align}
	where, $\Gamma_i=\frac{q_i B}{\om_i}$ is cyclotron frequency. For quantitative estimation, we have approximated it as $\frac{1}{\tau_B}$ ($\frac{1}{-\tau_B}$) for positive (negative) charged particle or antiparticle. This leads to the vanishing Hall components at zero chemical potential.
	The conductivity equations derived above are equivalent to their frequency-dependent counterparts obtained in the Drude model. In Drude model~\cite{SMITH1968126},  real part of frequency-dependent conductivity is $\sigma(\om) = \frac{\sigma_0}{1+\om^2 \tau^2}$, where $\sigma_0$ is conductivity for a static electric field,  $\om$ is frequency of time dependent electric field, and $\tau$ is collision time. In Eq. (\ref{ele-cond}), we find similar dependency on frequency in the conductivities via $\chi_i (= \frac{\tau_R^i}{\tau_B} = \frac{\tau_R^i}{\tau_E})$, where the decay parameters $\tau_B$ or $\tau_E$ are inverse of frequency of time-varying field.

	For the case of a constant electric and magnetic field, we can derive the electrical conductivity components using the RTA formalism, and the expressions are (detailed derivation can be found in~\cite{PhysRevD.102.114015}),
	\begin{align}
	\sigma_e &= \frac{1}{3T} \sum_i g_i (q_{i})^2 \int \frac{d^3|\vk_i|}{(2\pi)^3}
	\frac{ \vk^2_i}{\om_i^2} f^0_i(1\mp f^0_i) \tau_R^i \frac{1}{1+(\tau_R^i \Gamma_i)^2},\nn\\
	\sigma_H &= \frac{1}{3T} \sum_i g_i (q_{i})^2 \int \frac{d^3|\vk_i|}{(2\pi)^3}
	\frac{ \vk^2_i}{\om_i^2} f^0_i(1\mp f^0_i) \tau_R^i \frac{\tau_R^i \Gamma_i}{1+(\tau_R^i \Gamma_i)^2},
	\label{elec-cons}
	\end{align}
	In the absence of a magnetic field ($B = 0$), the Hall component vanishes, and the Ohmic conductivity becomes
	\begin{equation}
	\sigma_e = \frac{1}{3T} \sum_i g_i (q_{i})^2 \int \frac{d^3|\vk_i|}{(2\pi)^3}
	\frac{ \vk^2_i}{\om_i^2} f^0_i(1\mp f^0_i)~ \tau_R^i~.
	\label{elc-B0}
	\end{equation}

	\subsection{Thermal conductivity}
	\label{subsec:TC}
	The temperature gradient in a system is equilibrated by the flow of heat, and the corresponding current is determined by a coefficient called thermal conductivity. The thermal conductivity of a charged fluid becomes anisotropic in the presence of a magnetic field; instead of a single coefficient of thermal conductivity, we get multi-component thermal conductivity. In the presence of a time-varying magnetic field, heat current in the fluid rest frame can be expressed as~\cite{PhysRevD.106.034008}, 
	\begin{align}\label{I1}
	\vec{I} &= \kappa_0  \vec{\nabla}T +\bar{\kappa_1} (\vec{\nabla}T \times \vec{B}) +\bar{\kappa_2} (\vec{\nabla}T \times \dot{\vec{B}})\nonumber \\ 
	&= \kappa_0 \vec{\nabla}T +  (\kappa_1+\kappa_2) (\vec{\nabla}T \times \hat{b})\nonumber \\ 
	&= \kappa_0 \vec{\nabla}T +  \kappa_H (\vec{\nabla}T \times \hat{b})~.   
	\end{align}
	$\kappa_0$ is the leading component of thermal conductivity along the temperature gradient, and the Hall components are $\kappa_1=\bar{\kappa_1} {B}$, $\kappa_2 =\bar{\kappa_2} \dot{{B}}$.
	
	Now, the fluid property of a medium can be described by the energy-momentum tensor $T^{\mu\nu}$, particle four flow $N^{\mu}$, and with their conservation laws. In the kinetic theory, these quantities can be expressed in terms of the particle's energy, momentum, and phase space integration as
	\begin{align}\label{1.1}
	T^{\mu\nu}=\sum_{i}g_i\int{\frac{d^3|\vk_i|}{(2\pi)^3}\,\frac{{k}_i^{\mu} {k}_i^{\nu}}{\om_i}\,f_i},
	\nn\\
	N^{\mu}=\sum_{i}g_i\int{\frac{d^3|\vk_i|}{(2\pi)^3}\,\frac{{k}_i^{\mu}}{\om_i}\,f_i}.
	\end{align}
	Where, particle four momentum of $\textit{i}$th species is defined as $k^\mu_i = (\om_i, \vec{k_i})$.
	$T^{\mu\nu}$ and $N^{\mu}$ can be expressed as the sum of the ideal and dissipative part respectively as
	\begin{align}
	T^{\mu\nu} = T^{\mu\nu}_{\rm ideal} + \delta T^{\mu\nu}, \nn\\
	N^{\mu} = N^{\mu}_{\rm ideal} + \delta N^{\mu}.
	\end{align}
	The ideal parts correspond to the equilibrium distribution function $f_0$, and the dissipative part corresponds to deviated part of the distribution function $\delta f$.
	Here the conserved charge is baryon number, b$_i$. So, the heat current is defined as the energy flow related to the baryon current. 
		Now, in the first-order hydrodynamics~\cite{DeGroot:1980dk}, heat current is proportional to
		the gradient of the thermal potential as
		\begin{align}
		I^{\mu}&= -\kappa \frac{nT^2}{\epsilon+p}\nabla^{\mu} \Big(\frac{\mu}{T}\Big),	
		\end{align}
		where $\kappa$ represents the thermal conductivity and $\nabla^{\mu}=\partial^\mu-u^{\mu}u^{\nu}\partial_\nu$.  To derive the expression of thermal conductivity, we work in a local rest frame (LRF) of fluid, where fluid four velocity $u^\mu \equiv (1, \Vec{0})$. In the LRF using the Gibbs-Duhem relation, the heat current three-vector takes the form
		\begin{align}
		\vec{I} = -\kappa \vec{\nabla}T~.
		\end{align}
		We can choose the velocity frame as the Landau or Eckart frames. However, the Landau matching condition is necessary to work with the relaxation time approximation\cite{PhysRevD.104.096016}. In the Landau frame, $T^{0j} = 0$, and in the Eckart frame, $N^{j} = 0$. However, in both the frame, the heat current is defined as~\cite{HOSOYA1985666, GAVIN1985826}
		\begin{align}\label{1.4}
		{ I^j}= \delta T^{0j} -h ~\delta N^j,  
		\end{align}
	where $h=\frac{\ep + P}{n}$ is the enthalpy per particle, $\ep$, $P$, and $n$ are total energy density, total pressure, and net baryon density of the system, respectively. 
	Employing Eq.~(\ref{1.1}) in Eq.~(\ref{1.4}), we can express the three vector form of heat current for $\textit{i}$th species particles in terms of microscopic quantities as 
	\begin{align}\label{1.5a}
	{\vec {I}}_i = \int \frac{d^3|\vk_i|}{(2\pi)^3} \frac{\vec {k}_i}{\om_i} (\om_i -{\rm b}_i h)\delta f_i.
	\end{align}
	To find $\delta f$, we solve the BTE in the presence of an external magnetic field under the RTA in the LRF
	\begin{align}\label{BTE-RTA-Th}
	\frac{\del f_i}{\del t} + \frac{\vk_i}{\om_i} \cdot \frac{\del f_i}{\del \vec{x}} + q_i \left(\frac{\vk_i}{\om_i} \times \vec{B}\right) \cdot \frac{\del f_i}{\del \vec{k_i}}
	= -\frac{\delta f_i}{\tau^i_R}~.
	\end{align}
	We can assume an ansatz of $\delta f_i$ for thermal conductivity as
	\begin{equation}\label{1.6}
	\delta f_i=({\vec{k_i}}.{\vec \Omega}_\kappa ) \frac{\partial f^0_i}{\partial \om_i}, 
	\end{equation} 
	where a general form of ${\vec{\Omega}_\kappa}$ up to first order time derivative of $\vec B$ can be expressed as
	\begin{align}\label{1.7}
	\vec{\Omega}_\kappa = \alpha_1\vec{B}+ \alpha_2\vec{\nabla}T+ \alpha_3(\vec{\nabla}T \times \vec{B})+\alpha_4 \dot{\vec{B}} +\alpha_5(\vec{\nabla}T \times \dot{\vec{B}}).
	\end{align} 
	Considering the magnetic field profile same as Eq.~(\ref{Mag-Field1}), we can find the unknown coefficients $\alpha_{i}$ ($i=(1, 2,.., 5)$) by solving Eq.~(\ref{BTE-RTA-Th}) employing Eq.~(\ref{1.7}) in Eq.~(\ref{1.6}). From Eqs.~(\ref{I1}) and (\ref{1.5a}), we can obtain the expressions of thermal conductivity components (see Appendix \ref{appendix2})
	\begin{widetext}
		\begin{align}\label{18}
		&\kappa_0 =  \frac{1}{3T^2} \sum_i g_i \int \frac{d^3|\vk_i|}{(2\pi)^3}\frac{\vec{k}^2_i}{\om_i^2}(\om_i - {\rm b}_i h)^2 f^0_i(1\mp f^0_i) \tau_R^i ~\frac{1}{(1+\chi_i + \chi_i^2)}~,\nn\\
		&{\kappa}_1 = \frac{1}{3T^2} \sum_i g_i \int \frac{d^3|\vk_i|}{(2\pi)^3}\frac{ \vec{k}^2_i}{\om_i^2}(\om_i - {\rm b}_i h)^2 f^0_i(1\mp f^0_i) \tau_R^i ~\frac{\chi_i}{(1+\chi_i)(1+\chi_i + \chi_i^2)} ~, \nn\\
		&{\kappa}_2 = \frac{1}{3T^2} \sum_i g_i \int \frac{d^3|\vk_i|}{(2\pi)^3}\frac{\vec {k}^2_i}{\om_i^2}(\om_i - {\rm b}_i h)^2 f^0_i(1\mp f^0_i) \tau_R^i ~\frac{\chi_i^2}{(1+\chi_i)(1+\chi_i + \chi_i^2)} ~, \nn\\
		{\rm or,}\nn\\
		&{\kappa}_H = \kappa_1 + \kappa_2 = \frac{1}{3T^2} \sum_i g_i \int \frac{d^3|\vk_i|}{(2\pi)^3}\frac{\vec {k}^2_i}{\om_i^2}(\om_i - {\rm b}_i h)^2 f^0_i(1\mp f^0_i) \tau_R^i ~\frac{\chi_i}{(1+\chi_i + \chi_i^2)} ~.
		\end{align}
	\end{widetext}
	Here, the approximation $\tau_B = \frac{\om_i}{q_iB}$ creates a discrepancy in the Hall components. The Hall coefficients depend on the sign of the particle's charge. Therefore, to generate the results, we use the minus (plus) sign in $\kappa_1$ and $\kappa_2$ (or $\kappa_H$) for negatively (positively) charged particles and antiparticles.
	
	Now, for the case of a constant magnetic field, we can derive the expression of thermal conductivity components using the RTA formalism (detailed calculation can be found in Ref.~\cite{PhysRevD.100.114004}):
	\begin{align}
	\kappa_0 = \frac{1}{3T^2} \sum_i g_i \int \frac{d^3|\vk_i|}{(2\pi)^3}\frac{\vec{k}^2_i}{\om_i^2}(\om_i - {\rm b}_i h)^2 \tau_R^i ~\frac{1}{1+(\tau_R^i \Gamma_i)^2}& \nn\\ 
	\times f^0_i(1\mp f^0_i)&,\nn\\
	\kappa_H = \frac{1}{3T^2} \sum_i g_i \int \frac{d^3|\vk_i|}{(2\pi)^3}\frac{\vec{k}^2_i}{\om_i^2}(\om_i - {\rm b}_i h)^2 
	\tau_R^i ~\frac{\tau_R^i \Gamma_i}{1+(\tau_R^i \Gamma_i)^2}& \nn\\ 
	\times f^0_i(1\mp f^0_i)~.&
	\label{th-cons}
	\end{align}
	In the absence of a magnetic field ($B = 0$), the Hall component vanishes, and conductivity becomes isotropic with
	\begin{equation}
	\kappa_0 = \frac{1}{3T^2} \sum_i g_i \int \frac{d^3|\vk_i|}{(2\pi)^3}\frac{\vec{k}^2_i}{\om_i^2}(\om_i - {\rm b}_i h)^2 f^0_i(1\mp f^0_i)~ \tau_R^i ~.
	\label{th-B0}
	\end{equation}
	
	\subsection{Electrical and thermal conductivity under the hadron resonance gas model}
	\label{HRG}
	The HRG model successfully explains the hadron yield and various results of lattice QCD in the hadronic temperature zone~\cite{borsanyi2012fluctuations, PhysRevD.86.034509, PhysRevLett.111.202302, PhysRevD.92.114505}. In the ideal HRG model, the system is considered a grand canonical ensemble of non-interacting particles. All the thermodynamical quantities can be found from the grand canonical potential.
	According to the ideal HRG model, any transport coefficient of a system would be equal to the sum of contributions from all hadron species. In the case of electrical conductivity, only the charge hadrons (baryons and mesons) will contribute. Therefore, components of electrical conductivity under the HRG model for three cases: zero magnetic fields ($B = 0$), constant electromagnetic field, and time-varying electromagnetic field ($B(t)$) would be
	as follows. For $B = 0$ case, from Eq.~(\ref{elc-B0})
	\begin{align}
	\sigma_e &= \frac{1}{3T} \sum_{\rm baryon} g_i (q_{i})^2 \int \frac{d^3|\vk_i|}{(2\pi)^3}
	\frac{ \vk^2_i}{\om_i^2} f^0_i(1 - f^0_i)~ \tau_R^i\nn\\
	& ~ + \frac{1}{3T} \sum_{\rm meson} g_i (q_{i})^2 \int \frac{d^3|\vk_i|}{(2\pi)^3}
	\frac{ \vk^2_i}{\om_i^2} f^0_i(1 + f^0_i)~ \tau_R^i~.
	\label{elec-B0_HRG}
	\end{align}
	For a constant electromagnetic field case, from Eq.~(\ref{elec-cons})
	\begin{widetext}
		\begin{align}\label{ele-cons_HRG}
		\sigma_e = \frac{1}{3T} \sum_{\rm baryon} g_i (q_{i})^2 \int \frac{d^3|\vk_i|}{(2\pi)^3}
		\frac{ \vk^2_i}{\om_i^2} f^0_i(1 - f^0_i) \tau_R^i \frac{1}{1+(\tau_R^i \Gamma_i)^2} + 
		\frac{1}{3T} \sum_{\rm meson} g_i (q_{i})^2 \int \frac{d^3|\vk_i|}{(2\pi)^3}
		\frac{ \vk^2_i}{\om_i^2} f^0_i(1 + f^0_i) \tau_R^i \frac{1}{1+(\tau_R^i \Gamma_i)^2},\nn\\
		\sigma_H = \frac{1}{3T} \sum_{\rm baryon} g_i (q_{i})^2 \int \frac{d^3|\vk_i|}{(2\pi)^3}
		\frac{ \vk^2_i}{\om_i^2} f^0_i(1 - f^0_i) \tau_R^i \frac{\tau_R^i \Gamma_i}{1+(\tau_R^i \Gamma_i)^2}
		+    
		\frac{1}{3T} \sum_{\rm meson} g_i (q_{i})^2 \int \frac{d^3|\vk_i|}{(2\pi)^3}
		\frac{ \vk^2_i}{\om_i^2} f^0_i(1 + f^0_i) \tau_R^i \frac{\tau_R^i \Gamma_i}{1+(\tau_R^i \Gamma_i)^2}.
		\end{align}
	\end{widetext}
	For the case of time-varying fields ($B(t)$), from Eqs.(\ref{C-EC}) and (\ref{ele-cond}) 
	\begin{widetext}
		\begin{align}
		&\sigma_e = \frac{1}{3T} \sum_{\rm baryon} g_i (q_{i})^2 \int \frac{d^3|\vk_i|}{(2\pi)^3}
		\frac{ \vk^2_i}{\om_i^2} f^0_i(1-f^0_i) \tau_R^i~ \frac{2 \chi_i^2+2 \chi_i+1}{(\chi_i+1) \left(\chi_i^2+1\right) \left(\chi_i^2+\chi_i+1\right)} \nn\\
		& ~~~~~+ 
		\frac{1}{3T} \sum_{\rm meson} g_i (q_{i})^2 \int \frac{d^3|\vk_i|}{(2\pi)^3}
		\frac{ \vk^2_i}{\om_i^2} f^0_i(1+f^0_i) \tau_R^i~ \frac{2 \chi_i^2+2 \chi_i+1}{(\chi_i+1) \left(\chi_i^2+1\right) \left(\chi_i^2+\chi_i+1\right)},\nn\\
		&\sigma_H = \frac{1}{3T} \sum_{\rm baryon} g_i (q_{i})^2 \int \frac{d^3|\vk_i|}{(2\pi)^3}
		\frac{ \vk^2_i}{\om_i^2} f^0_i(1-f^0_i) \tau_R^i~ \frac{\chi_i \left(\chi_i^3+\chi_i^2+2 \chi_i+1\right)}{(\chi_i+1) \left(\chi_i^2+1\right) \left(\chi_i^2+\chi_i+1\right)}  \nn\\
		&~~~~~+ 
		\frac{1}{3T} \sum_{\rm meson} g_i (q_{i})^2 \int \frac{d^3|\vk_i|}{(2\pi)^3}
		\frac{ \vk^2_i}{\om_i^2} f^0_i(1+f^0_i) \tau_R^i~ \frac{\chi_i \left(\chi_i^3+\chi_i^2+2 \chi_i+1\right)}{(\chi_i+1) \left(\chi_i^2+1\right) \left(\chi_i^2+\chi_i+1\right)}~.
		\label{Work_Elec}
		\end{align}
	\end{widetext}
	Note that in $B(t)$ case, for simplicity we have considered $\tau_E = \tau_B$ and $\chi_i = \frac{\tau_R^i}{\tau_B} = \frac{\tau_R^i}{\tau_E}$. Also, Hall conductivity is negative (positive) for negatively (positively) charged particles and antiparticles.

	Now, according to the definition of heat flow, only the baryon will contribute to thermal conductivity. Unlike electrical conductivity, neutral baryons will also contribute to the Ohmic component of thermal conductivity. So, the thermal conductivity components under the HRG model can be expressed for the three cases as follows.
	For $B = 0$ case, from Eq.~(\ref{th-B0})
	\begin{equation}
	\kappa_0 = \frac{1}{3T^2} \sum_{\rm baryon} g_i \int \frac{d^3|\vk_i|}{(2\pi)^3}\frac{\vec{k}^2_i}{\om_i^2}(\om_i - {\rm b}_i h)^2 f^0_i(1 - f^0_i)~ \tau_R^i ~.
	\label{th-B0_HRG}
	\end{equation}
	For the constant magnetic field case, from Eq.~(\ref{th-cons})
	\begin{align}
	&\kappa_0 = \frac{1}{3T^2} \sum_{\rm baryon} g_i \int \frac{d^3|\vk_i|}{(2\pi)^3}\frac{\vec{k}^2_i}{\om_i^2}(\om_i - {\rm b}_i h)^2 f^0_i(1 - f^0_i) \tau_R^i  \nn\\ 
	&~~~~~~~~~~~~~~~~~~~~~~~~~~~~~~~~~~~~~~~~ \times \frac{1}{1+(\tau_R^i \Gamma_i)^2},\nn\\
	&\kappa_H = \frac{1}{3T^2} \sum_{\rm baryon} g_i \int \frac{d^3|\vk_i|}{(2\pi)^3}\frac{\vec{k}^2_i}{\om_i^2}(\om_i - {\rm b}_i h)^2 
	f^0_i(1 - f^0_i) \tau_R^i  \nn\\ 
	&~~~~~~~~~~~~~~~~~~~~~~~~~~~~~~~~~~~~~~~~ \times \frac{\tau_R^i \Gamma_i}{1+(\tau_R^i \Gamma_i)^2}~.
	\label{th-cons_HRG}
	\end{align}
	For $B(t)$ case, from Eq.~(\ref{18})
	\begin{align}
	&\kappa_0 =  \frac{1}{3T^2} \sum_{\rm baryon} g_i \int \frac{d^3|\vk_i|}{(2\pi)^3}\frac{\vec{k}^2_i}{\om_i^2}(\om_i - {\rm b}_i h)^2 f^0_i(1 - f^0_i) \tau_R^i  \nn\\ 
	&~~~~~~~~~~~~~~~~~~~~~~~~~~~~~~~~~~~~~~~~ \times\frac{1}{(1+\chi_i + \chi_i^2)}~,\nn\\
	&{\kappa}_H = \frac{1}{3T^2} \sum_{\rm baryon} g_i \int \frac{d^3|\vk_i|}{(2\pi)^3}\frac{\vec {k}^2_i}{\om_i^2}(\om_i - {\rm b}_i h)^2 f^0_i(1 - f^0_i) \tau_R^i \nn\\ 
	&~~~~~~~~~~~~~~~~~~~~~~~~~~~~~~~~~~~~~~~~ \times\frac{\chi_i}{(1+\chi_i + \chi_i^2)}~.
	\label{Work_Ther}
	\end{align}
	Here, b$_i$ is the baryon quantum number of $\textit{i}$th species. Enthalpy per particle, $h=\frac{\rm total ~enthalpy~ of~ the~ system}{\rm Net~ baryon~ density}$. Eqs.~(\ref{elec-B0_HRG}) to (\ref{Work_Ther}) are the working formula for the results obtained in the next section.

	\subsection{The relaxation time under the HRG Model}
	\label{RT}
	In the case of the ideal HRG model, particles are noninteracting. Here we will calculate the relaxation time of a hadron for $2 \rightarrow 2$ elastic collision process with all the other hadrons (baryons and mesons) that exist in the medium. Relaxation time $\tau_R$ is energy (or momentum) dependent. However, a momentum-independent relaxation time is required for RTA to obey the energy-momentum conservation law. Therefore, we calculate the thermal average relaxation time for the process mentioned above between $i^{\rm th}$ and $j^{\rm th}$ particles. The thermal average relaxation time of particle $i$ can be expressed as~\cite{GONDOLO1991145, PhysRevD.94.094002, PhysRevD.100.114004}
	\begin{equation}
	{(\tau_R^i)}^{-1}=\sum_{j} n_j \langle\sigma_{ij}v_{ij}\rangle~.
	\label{Eq-relax}
	\end{equation}
	Where, $j$ runs for all the hadrons in the medium except $i$; $n_j=\int\frac{d^3\vk_j}{(2\pi)^3}f_j^0$, and thermal averaged cross-section $\langle \sigma_{ij}v_{ij}\rangle$ in the center of mass frame can be expressed as
	\begin{align}
	\langle\sigma_{ij}v_{ij}\rangle = \frac{\sigma}{8Tm_i^2m_j^2K_2(m_i/T)K_2(m_j/T)}\int_{(m_i+m_j)^2}^{\infty}ds\times\nn\\ \frac{[s-(m_i-m_j)^2]}{\sqrt{s}}
	\times [s-(m_i+m_j)^2]K_1(\sqrt{s}/T)~.
	\label{equ73}
	\end{align}
	Here, $\sqrt{s}$ is the center of mass energy; $m_i$, $m_j$ are the mass of the particles; $K_1$ and $K_2$ are the modified Bessel functions of the second kind of first and second order, respectively. $\sigma = \pi (r_i + r_j)^2$ is total scattering cross-section for hard sphere of radius $r_i$ and $r_j$. For quantitative estimations, we considered the same radius for all mesons as $0.2~$fm, and for all baryons, it is $0.62~$fm~\cite{PhysRevC.98.014907}.
	
	\section{Result}
	\label{sec-result}
	For quantitative estimation of conductivities, we used the ideal HRG model, taking into account all the resonances (baryons and mesons) from the particle data group (PDG)~\cite{10.1093/ptep/ptac097} having spin $0, 1, 1/2$, and $3/2$. 
	In electrical conductivity, only the charged hadrons contribute. In thermal conductivity, only baryons (charged and neutral) directly contribute, and mesons contribute via the particle's relaxation time and enthalpy of the system. 
	
	\begin{figure}
		\centering
		\includegraphics[scale=0.35]{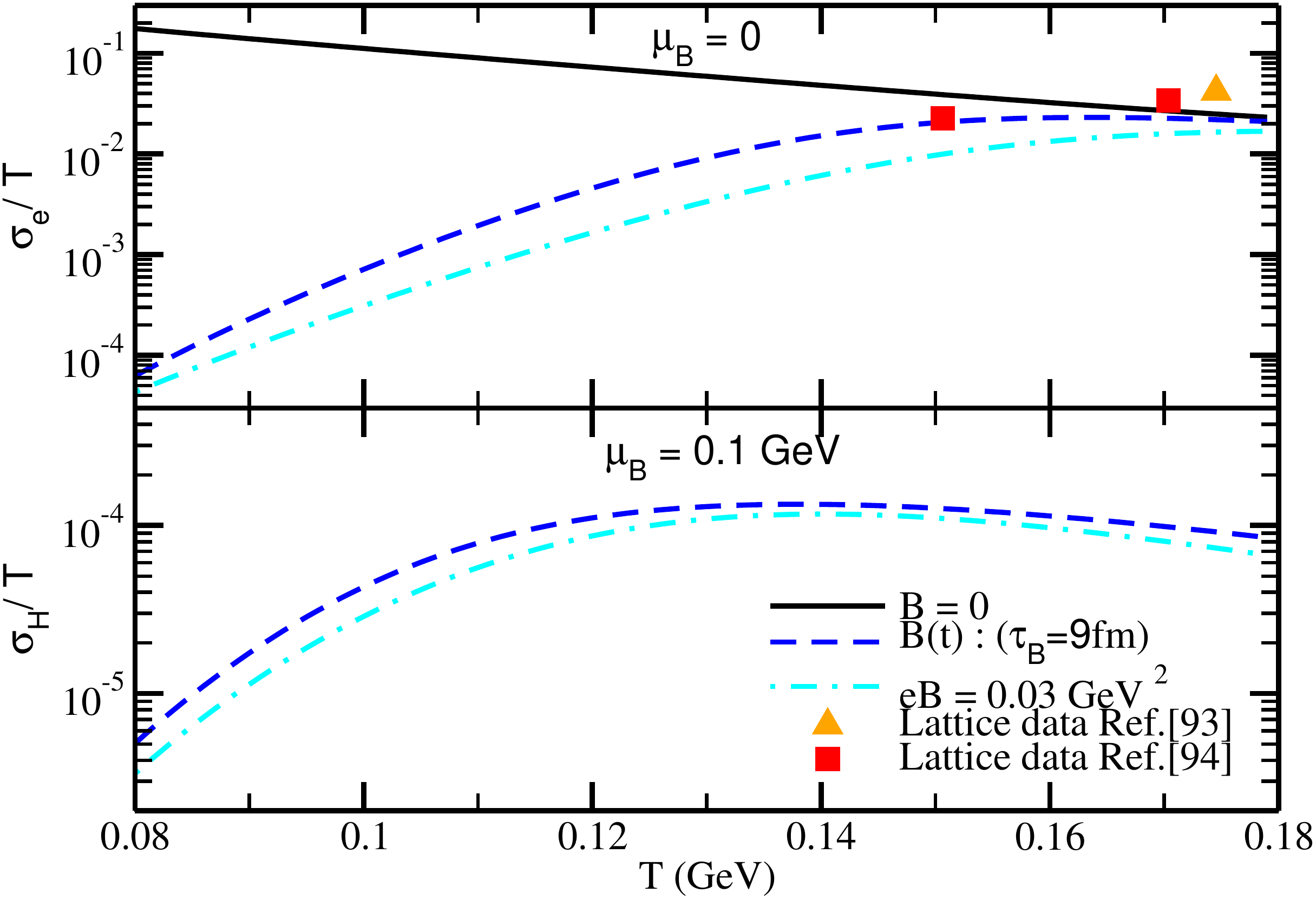}
		\caption{Scaled electrical conductivity components ($\sigma_e,~\sigma_H/T$) as a function of temperature ($T$). The black solid line for zero magnetic fields ($B = 0$), blue dashed line for time-varying fields (B(t)), and cyan dash-dotted line for constant magnetic field $eB = 0.03~$GeV$^2$. Lattice results of $\sigma_e/T$ from Refs.~\cite{Ding:2016hua, Amato:2013naa} are also shown for comparison.}
		\label{Fig-elec}
	\end{figure}
	In Fig.~(\ref{Fig-elec}), we have plotted the Ohmic (upper panel) and Hall (lower panel) components of scaled electrical conductivity as a function of $T$. For Ohmic component $\sigma_e$, we have shown results at zero baryon chemical potential for three cases: zero magnetic field ($B = 0$), the constant magnetic field with $eB = 0.03$~GeV$^2$, and time-varying fields ($B(t)$) with a fixed decay parameter $\tau_B = 9~$fm. For $B = 0$ case, $\sigma_e$ decrease with $T$. At $B = 0$, $\sigma_e$ is directly proportional to $\tau_R$. Therefore, the major contribution in the $T$ dependent profile is from relaxation time $\tau_R$, which decreases with increasing $T$ in the HRG system with a fixed scattering cross-section. For quantitative estimation of the relaxation time, we fixed baryons' radius $0.62~$fm and mesons' radius $0.2~$fm~\cite{PhysRevC.98.014907}. For the case of $B(t)$ and $eB = 0.03~$GeV$^2$, $\sigma_e$ increase with $T$ and saturates below $B = 0$ case at high $T$ in the hadronic temperature zone. This is because, unlike $B = 0$ case, $\sigma_e$ in the presence of a magnetic field is approximately inversely proportional to $\tau_R$ that can be found from Eq.~(\ref{Work_Elec}). 
	$\sigma_e$ for time-varying case with $\tau_B = 9~$fm remain higher than constant field case with $eB = 0.03~$GeV$^2$.
	This is because, in the $B(t)$ case, there is an additional contribution from $\sigma_e^{(1)}$ in Eq.~(\ref{ele-cond}) arising due to time-varying electric field. $T$ dependency for slowly varying field case is approximately the same as constant magnetic field case. Note that, in the case of a slowly-varying field, conductivity is independent of the magnitude of the field. However, at the constant magnetic field, conductivity depends on the magnitude of the field through cyclotron frequency. 
	We have compared our $B=0$ curve with lattice results of $\sigma_e/T$ from Refs.~\cite{Ding:2016hua, Amato:2013naa}, which reflects that the order of magnitude of our HRG estimations is a good agreement with earlier knowledge. It indirectly indicates that our tuning parameters for meson and baryon scattering cross-section or hard sphere radius have been well guessed. Although, a broad numerical band of estimated electrical conductivity(see the Table of Ref.~\cite{SG_el_PRD}) of quark and hadronic matter can be found in the literature, which is beyond the scope of the present work. 
	
	In the lower panel of Fig.~(\ref{Fig-elec}), the Hall component of electrical conductivity is plotted for time-varying electromagnetic field and constant magnetic field case. Hall conductivity only exists at a finite chemical potential. At zero $\mu_B$, Hall conductivity vanishes due to equal and opposite contributions from particle and antiparticle. There is a sign dependency on the particle's charge in the expression of $\sigma_H$ in Eq.~(\ref{Work_Elec}). Therefore, the absolute value of $\sigma_H$ is minimal compared to $\sigma_e$ for the HRG system. 
	$\sigma_H$ for $B(t)$ case is higher than constant $eB$ case because of extra contributions from $\sigma^{(1)}_H$ and $\sigma^{(2)}_H$ due to time time-varying electric and magnetic field, respectively.
	
	\begin{figure}
		\centering
		\includegraphics[scale=0.35]{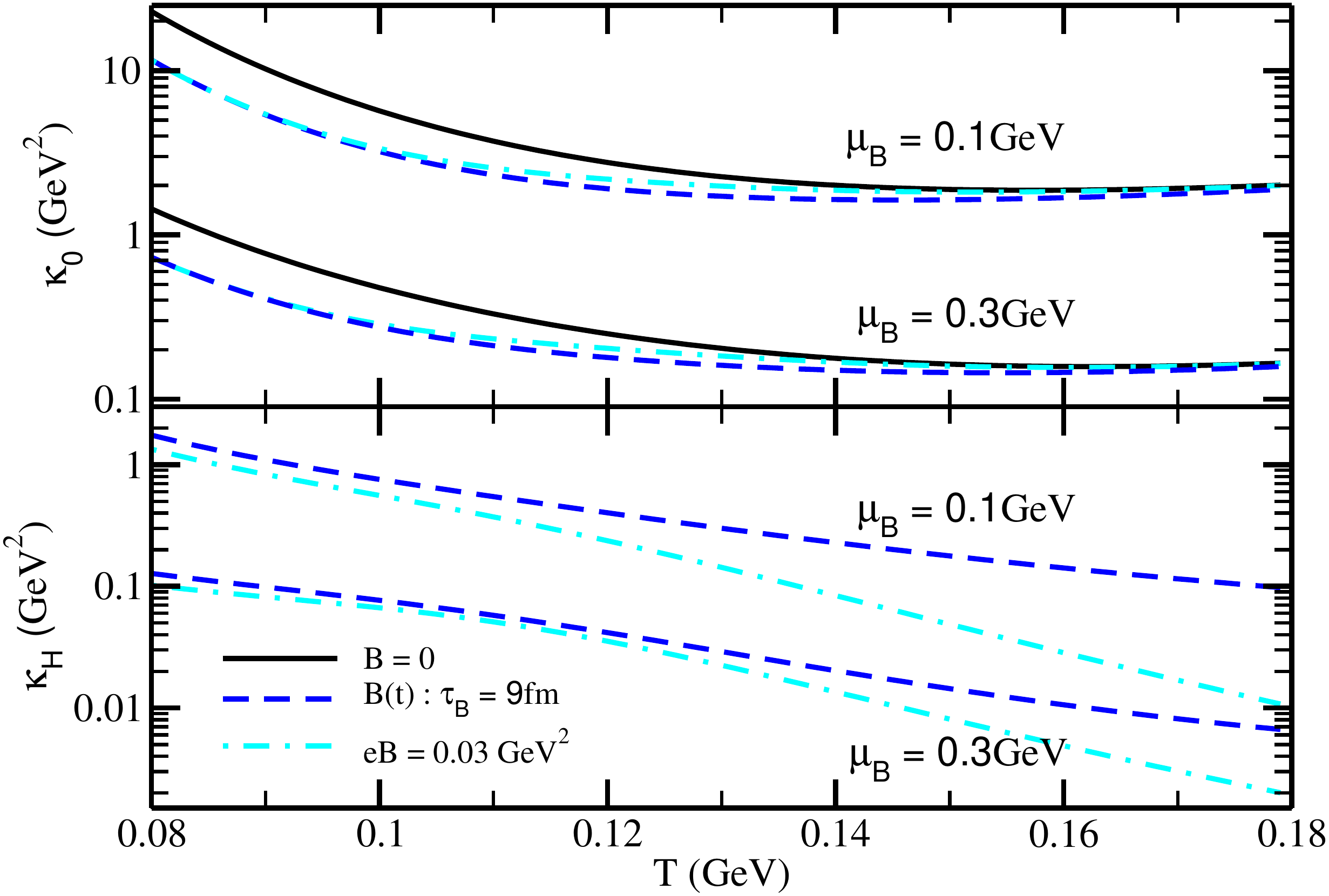}
		\caption{Components of thermal conductivity ($\kappa$) as a function of temperature ($T$).}
		\label{Fig-therm}
	\end{figure}
	In Fig.~(\ref{Fig-therm}), we have plotted thermal conductivity components as a function of temperature at baryon chemical potential $\mu_B = 0.1,~ 0.3~$GeV. In the upper panel, Ohmic-type conductivity $\kappa_0$ is plotted for three cases: zero magnetic field ($B = 0$), the constant magnetic field with $eB = 0.03$~GeV$^2$, and time-varying fields $B(t)$ with a fixed decay parameter $\tau_B = 9~$fm. $\kappa_0$ decrease with $T$ for all the three cases. For $B = 0$ case, $\kappa_0$ is proportional to $\tau_R$ which explain the $T$ dependency. For the case of $B(t)$ and $eB = 0.03~$GeV$^2$, $\kappa_0$ is inversely proportional to $\tau_R$, but its trend can not be observed in the curves. It indicates some other sources which control the temperature profile. Unlike electrical conductivity, thermal conductivity integrand carries enthalpy density per net baryon density, which blows up at low temperatures. This is one of the major reasons for getting the decreasing profile of $\kappa_0(T)$. Another source is neutral hadrons, which also contribute to thermal conductivity. Neutral particles do not get affected by the electromagnetic field or Lorentz force. Therefore, $\kappa_0$ for neutral particles is the same as for the $B = 0$ case, which decreases with temperature. 
	With the increase in $\mu_B$, the thermal conductivity decreases because the rise in baryon chemical potential increases the enthalpy per particle with $T$, which reduces the thermal conductivity. Similar results were also found in Ref.~\cite{PhysRevD.100.114004}. An interesting result is that $\kappa_0$ of $B(t)$ is a bit smaller than that of constant $B$, which can be understood by comparing their expressions.      
	
\begin{figure}
		\centering
		\includegraphics[scale=0.35]{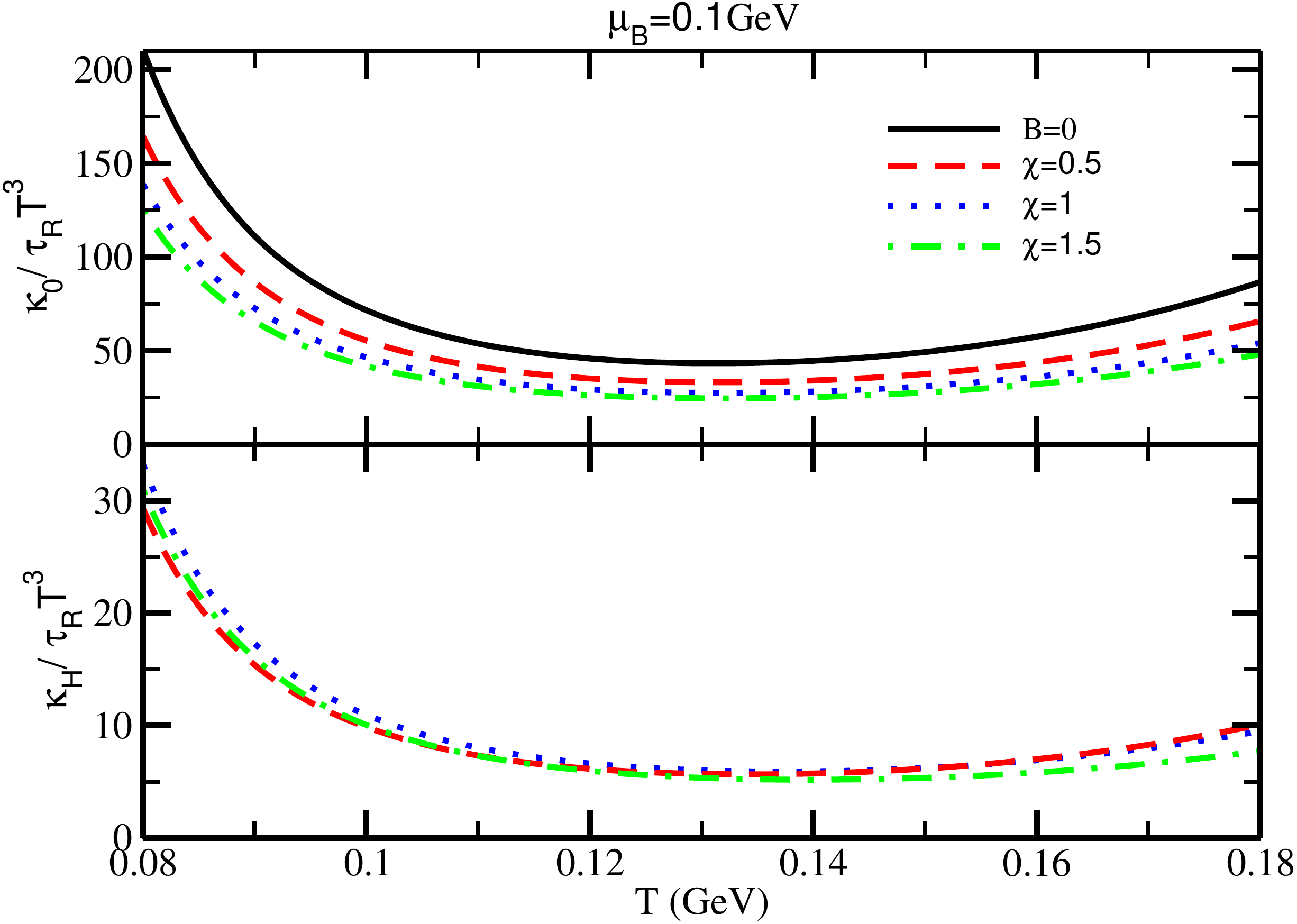}
		\caption{Scaled thermal conductivity as a function of temperature. Here, $\chi=\frac{\tau_R}{\tau_B}$. A higher value of $\chi$ represents faster decay of the field. The faster the decay, the lower the conductivity.}
		\label{fig:RatioTh}
	\end{figure}

	In the lower panel of Fig.~(\ref{Fig-therm}), the Hall component $\kappa_H$ is plotted for $B(t)$ and $eB = 0.03~$GeV$^2$ cases. $\kappa_H$ decreases with $T$ for both the cases. Although $\kappa_H$ is approximately inversely proportional to $\tau_R$, $T$ dependency is dominated by the thermodynamical phase-space part, carrying enthalpy density per net baryon density, which decreases with $T$. 
	This will be better understood when we discuss Fig.~(\ref{fig:RatioTh}) in the coming paragraphs. 
	We notice that the magnitude of $\kappa_H$ is roughly one order less than $\kappa_0$, for two reasons. Neutral hadrons don't participate in Hall thermal conductivity, and for charged hadrons, anti-particle contribution is subtracted from particle contribution instead of addition. The value of $\kappa_H$ for $B(t)$ case is higher than the constant field case. Because, in the case of $B(t)$, $\kappa_H$ has an additional component $\kappa_2$ (Eq.~\ref{18}), which is missing in the constant magnetic field case (Eq.~\ref{th-cons}).

	In Fig.~(\ref{fig:RatioTh}), thermal conductivity components scaled by $\tau_R T^3$ are plotted against $T$ at $\mu_B=0.1$~GeV for the time-varying case. The upper and lower panels represent the $\kappa_0$ and $\kappa_H$ components. Here, we have plotted the results for three different ratios of relaxation time to decay parameter, $\chi(=\frac{\tau_R}{\tau_B,\tau_E})=0.5,~1,~1.5$. To focus on the effect of the decay parameter, we assume the same $\tau_R$ for all the hadrons. $\chi>1 {~\rm and} <1$ correspond to faster and slower decay of fields respectively. $\kappa_0$ for $B=0$ case is also shown for comparison. Although $\kappa_0$ decreases in the presence of a magnetic field, $\kappa_0$ increases with the decrease in $\chi$. This indicates that the slower the decay of fields higher the conductivity along the temperature gradient. This effect is irrespective of the magnitude of the fields. However, the Hall component shows different behavior. $\kappa_H$ is higher when the decay parameter equals relaxation time. Moreover, $\kappa_H$ has a relatively small dependency on the decay parameters. Here, $T$ dependency from $\tau_R$ is canceled due to scaling and fixed value of $\tau_R$. Therefore, temperature dependency only arises due to the thermodynamical phase-space part in the expression of $\kappa_0$ and $\kappa_H$, which decrease with $T$. The combined effect from the thermodynamical phase-space part and $\tau_R$ on $T$ profile of thermal conductivity have already been seen in Fig.~(\ref{Fig-therm}).

	\begin{figure}
		\centering
		\includegraphics[scale=0.35]{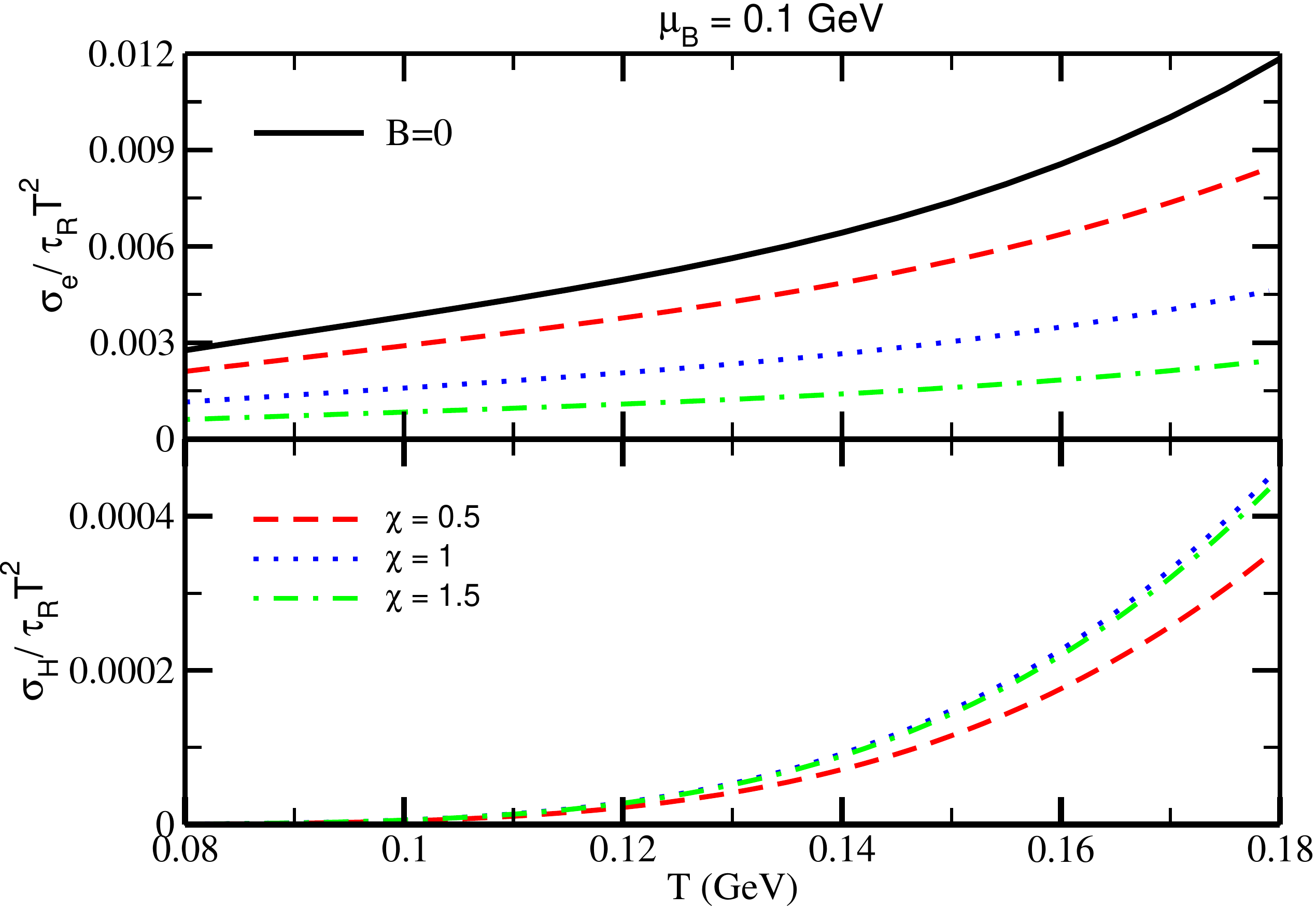}
		\caption{Scaled electrical conductivity as a function of temperature. Here, $\chi=\frac{\tau_R}{\tau_B}$. A higher value of $\chi$ represents faster decay of the field. Faster the decay, the lower the conductivity.}
		\label{fig:Ratio}
	\end{figure}
	In Fig.~(\ref{fig:Ratio}), components of electrical conductivity scaled with $\tau_R T^2$ are plotted against $T$. Electrical conductivity components also show a similar dependency on decay parameters as in the case of thermal conductivity in Fig.~\ref{fig:RatioTh}. 
	Here, unlike thermal conductivity, the thermodynamical phase-space part of $\sigma_e$ and $\sigma_H$ increase with $T$. 
	So, according to these Figs.~(\ref{fig:RatioTh}), (\ref{fig:Ratio}), we can roughly conclude that slower decay ($\chi < 1$) of the magnetic field can enhance the electrical and thermal conductivity along perpendicular direction of the magnetic field. Whereas Hall components of electrical and thermal conductivity become maximum when relaxation time and decay time become the same or $\chi = 1$. The aim of Figs.~(\ref{fig:RatioTh}), (\ref{fig:Ratio}) is just to see the role of the faster or slower decay time of magnetic field on the electrical and thermal conductivity, whereas the aim of Figs.~(\ref{Fig-elec}) and (\ref{Fig-therm}) were to provide the exact estimation by including $T$ dependency of relaxation time.

	\begin{figure}
		\centering
		\includegraphics[scale=0.35]{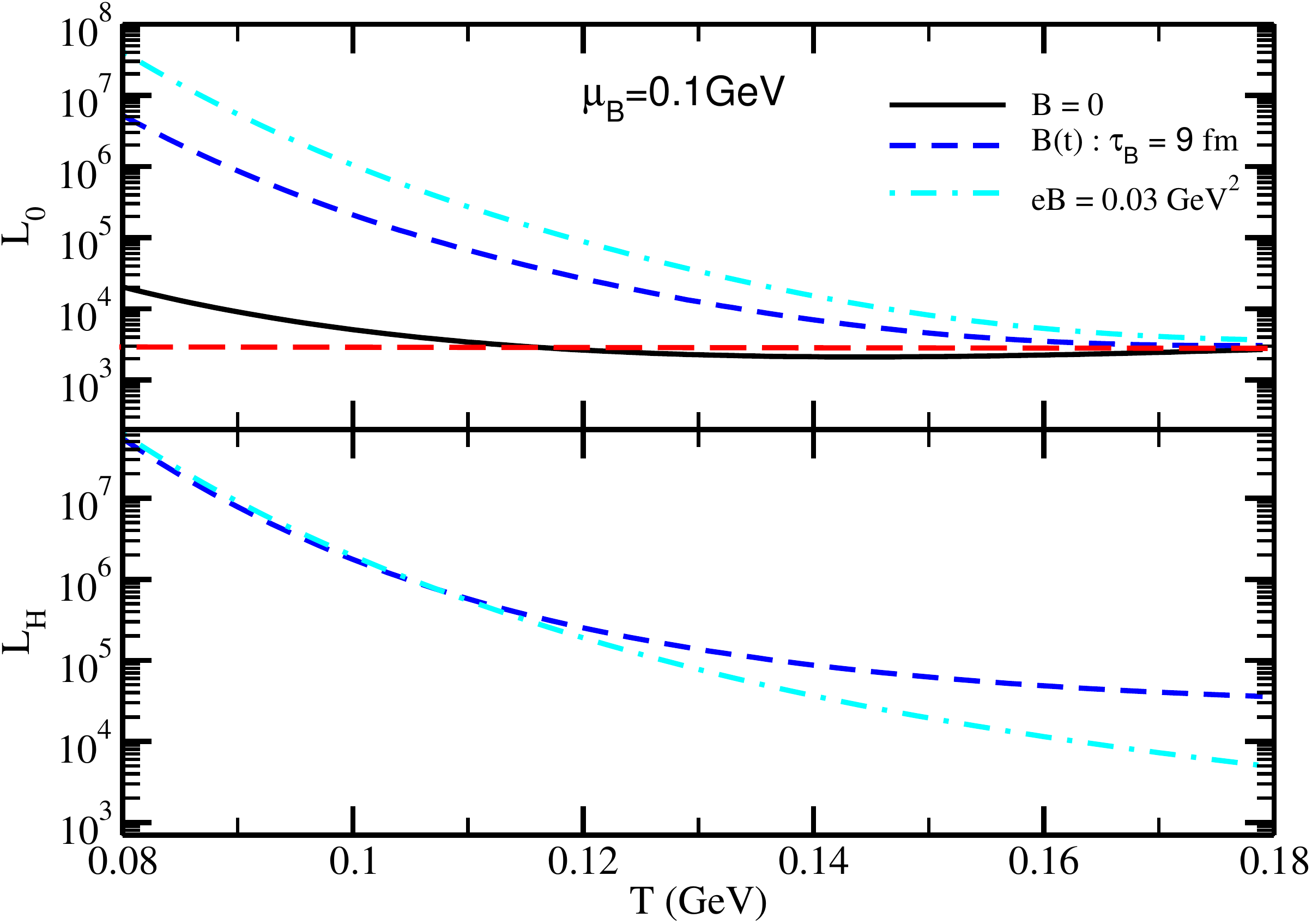}
		\caption{Lorenz number as a function of temperature. In the upper panel L$_0=\frac{\kappa_0}{T\times \sigma_e}$ and in the lower panel L$_H=\frac{\kappa_H}{T\times \sigma_H}$. Red horizontal straight line (long dashed) is drawn for reference to show a violation of the WF law.}
		\label{fig:L}
	\end{figure}
	In the end, we build the ratio of thermal and electrical conductivity components with the aim of the validity checking of the Wiedemann-Franz (WF) law. This ratio tells the interplay between electrical and thermal conductivity for any substance, which helps us to understand the relative importance of charge and heat diffusion in that substance. According to the WF law, in the case of metals, the thermal-to-electrical conductivity ratio is proportional to the system's temperature, and the proportionality constant is known as the Lorenz number. 
	The Lorenz number of quark and hadronic matter have been calculated in a few Refs.~\cite{PhysRevD.106.034008,Shaikh:2022sky,Pradhan:2022gbm,WF_RS1}, where most of them~\cite{PhysRevD.106.034008,Shaikh:2022sky,Pradhan:2022gbm} have found that it varies with temperature, indicating a violation of the  WF law. Refs.~\cite{PhysRevD.106.034008,Shaikh:2022sky} estimated Lorenz number in the presence of an electromagnetic field for the QGP medium using a quasiparticle-based model, and Ref.~\cite{Pradhan:2022gbm} estimated the same using the van der Waals HRG (VDWHRG)
	model in the absence of the magnetic field. These studies found that WF law is violated in QGP and HRG matters. In contrast, Ref.~\cite{WF_RS1} found that the Lorenz number is independent of temperature for the hot QCD matter under the color String Percolation (CSP) model, where only quarks participate in both the thermal and electrical conduction, which is analogous to the case of metal, where electrons participate dominantly in both the thermal and electrical conduction. Also, instead of the hydrodynamics-based expression for thermal conductivity, the expression obtained in Ref.~\cite{WF_RS1} is similar to the free electron theory in metal. This difference between metal and hydrodynamics-based expression of electronic conduction becomes a matter of interest in
	both experimental~~\cite{doi:10.1126/science.aad0343}, and theoretical~\cite{Lucas:2018kwo, Narozhny:2022ncn} condense matter physics, which opened a new research domain, called electron hydrodynamics~\cite{Narozhny:2022ncn}.

	In Fig.~(\ref{fig:L}), Lorenz number, L$_0 = \kappa_0/(\sigma_e T)$ (upper panel) and Hall-Lorenz number, L$_H = \kappa_H/(\sigma_H T)$ (lower) are plotted against $T$ axis at $\mu_B=0.1$~GeV for $B = 0$ case, time-varying case ($B(t)$) with $\tau_B=9$~fm, and constant magnetic field case with $eB=0.03$~GeV$^2$ for HRG matter. In the upper panel, we have found that the Lorenz number varies with temperature for the three cases, indicating a possible violation of the Wiedemann-Franz law in the HRG matter, for heat conductivity related to baryon current. 
	In the absence of a magnetic field, the Lorenz number varies slowly with temperature in contrast to the constant and time-dependent electromagnetic field. Although L$_0$ depends on $T$ throughout the whole hadronic temperature range, it behaves independently of $T$ at the high $T$ region, suggesting a slight metal-like behavior. A Red dashed straight horizontal line is drawn for reference from the saturated values of Lorenz number at high temperature, which indicates that $B=0$ and $B\neq 0$ cases can have different saturation points. If we focus on the deviation from these saturation values, we see the ranking - constant $B$ case $>$ $B(t)$ case $>$ $B=0$ case.
	In the HRG matter, we find another ingredient that can be responsible for violating WF law. Here, charged and neutral baryons contribute to thermal conductivity, while the charged baryons and mesons contribute to electrical conductivity. This in-equal component contribution in thermal and electrical conductivity can be responsible for the Lorenz number to be $T$ dependent, which was already found in earlier Ref.~\cite{Pradhan:2022gbm} for zero magnetic fields. 
	In the lower panel of Fig.~(\ref{fig:L}), we can find that the Hall-Lorentz number L$_H$, which is also deviating mostly from any saturation values. The fact shows similarity with the condensed matter physics, where Hall-Lorenz number for various metals and alloys also shows a violation of the WF law \cite{MATUSIAK2006376,PhysRevLett.100.016601}.

	\section{Summary and Conclusion} 
	\label{sec-summary}
	In summary, we have estimated the thermal and electrical conductivity in a time-varying electromagnetic field in the hadronic matter.
	We know thermal conductivity is the proportionality coefficient between thermal current density and temperature gradient,
	while electrical conductivity is the proportionality coefficient between electric current density and electric field.
	Following those macroscopic definitions, we have calculated their microscopic expressions in the kinetic theory framework by using 
	the Boltzmann transport equation under the relaxation time approximation. 
	In the presence of a magnetic field, the conductivity tensor becomes anisotropic. The two components
	- perpendicular and Hall components will depend on the magnetic field, but the parallel one remains independent of the magnetic field.
	If z is the magnetic field direction, then x or y directional conduction will be perpendicular components, x-y or y-x conduction
	will be Hall components, and z directional conduction will be parallel components. This anisotropic conduction picture can be found using a magnetic field and Lorentz force term
	in the Boltzmann equation. Now, when one goes to the (slowly) time-varying magnetic field,
	then along with the leading coefficients, coming for a constant magnetic field, some additional coefficients will contribute to the calculations. 
	
	In the present work, we have estimated thermal and electrical conductivity and their ratio to check the validation of the so-called WF law for HRG system in the presence
	of a time-varying magnetic field, which is also compared with constant and zero magnetic field cases.
	Relaxation time for all baryons and mesons is fixed by two different values of hard-sphere scattering cross sections, and
	they follow a decreasing (roughly exponential) temperature profile. Using that relaxation time in electrical and thermal conductivity in the absence of a magnetic field,
	we get a similar kind of decreasing profile in the temperature axis. Now, the role of a constant magnetic field and time-varying magnetic field will be to introduce the new time scales - cyclotron time and decay time, respectively, which will couple with relaxation time and make an effectively shorter relaxation time in the system. As a result, the magnetic field always reduces the electrical and thermal conductivity along a perpendicular direction. We have found the ranking - zero magnetic field $>$ time-varying magnetic field $>$ constant magnetic field for electrical conductivity and ranking - zero magnetic field $>$ constant magnetic field $>$ time-varying magnetic field for thermal conductivity. While searching for the validity of WF law, we have found that the ratio between thermal and electrical conductivity follows a similar ranking of thermal conductivity due to its dominant magnitude. We have observed the deviation of WF law in the entire hadronic temperature domain, although, at high temperatures, a possibility of saturation value is noticed. Concerning the saturation value, if we measure the deviation of WF law, then we will find a ranking - constant magnetic field $>$ time-varying magnetic field $>$ zero magnetic field. We have also estimated the Hall-Lorenz number for the HRG matter, which also shows a possible violation of WF law in the presence of a magnetic field. It is important to note here that this study specifically applies to heat conductivity related to baryon current. HRG is a multi-component system with different conserved charges, baryon, (electric) charge, and strangeness. To have a complete understanding of the HRG system in the presence of a magnetic field, one needs to include thermal conductivity related to (electric) charge and strangeness current, whereas Refs.~\cite{PhysRevLett.120.242301, PhysRevD.101.076007} would be useful. Ref.~\cite{PhysRevD.101.076007} shows that at zero baryon chemical potential, the relation between electrical conductivity and charge diffusion coefficient is a manifestation of WF law.
		Furthermore, the relaxation time calculated here, considering hard sphere scattering with a fixed cross-section for all the baryons and mesons differs from the actual scenario. Also, results obtained using the RTA method would be deviated from the Chapman-Enskog method, as shown in Ref.~\cite{PhysRevD.101.076007}.       

	Some of the observations in the present study are as follows. 
	The transport coefficients for the time-varying fields are independent of the initial field strength in the weak field limit that is being considered here.  
	Additionally, we have noticed that for faster decay of the fields, the perpendicular components of electrical and thermal conductivity decrease when the decay time is shorter than the relaxation time.
	In other words, conduction can be enhanced in a slow decay picture. On the other hand, the Hall conductivity can be maximum in a specific case when relaxation time equals the decay time of the field.  
	
	Present work indicates that the actual time-varying magnetic field picture of transport coefficient calculations is a bit more tedious, but estimation can differ from the constant magnetic field picture. Other transport coefficients like shear viscosity, bulk viscosity, etc., can also be studied in the future for this time-varying magnetic field.
	The present study is restricted by the slowly time-varying framework, which may be considered an open and future scope to find a general framework. 
	
	
	\section*{Acknowledgement}
	KS acknowledges the doctoral fellowship from UGC, Government of India. JD and RS gratefully acknowledge the DAE-DST, Govt. of India funding under the mega-science project – “Indian participation in the ALICE experiment at CERN” bearing Project No. SR/MF/PS-02/2021-
	IITI (E-37123). This research work uses the computing facilities under the DST-FIST scheme (Grant No. SR/FST/PSI-225/2016) of the Department of Science and Technology (DST), Government of India.
	
	\appendix
	\section*{Appendices}
	\label{sec-appendix}
	\setcounter{equation}{0}
	\renewcommand{\thesubsection}{\Alph{subsection}}
	\renewcommand{\theequation}{\thesubsection.\arabic{equation}}
	\subsection{Appendix: Electrical Conductivity}
	\label{appendix1}
	In the presence of an external electromagnetic field, BTE under RTA can be expressed as
	\begin{align}\label{4.2}
	\frac{\del f_i}{\del t} + \frac{\vk_i}{\om_i} \cdot \frac{\del f_i}{\del \vec{x}} + q_i \left(\vec{E} + \frac{\vk_i}{\om_i} \times \vec{B}\right) \cdot \frac{\del f_i}{\del \vec{k_i}}
	= -\frac{\delta f_i}{\tau^i_R}~.
	\end{align}
	From Eq.~(\ref{delta-f}), we have,
	\begin{align}\label{adelta-f}
	\delta f_i = (\vec{k_i} \cdot \vec{\Omega}_\sigma) \frac{\del f^0_i}{\del \om}~.
	\end{align}  
	Substitution of Eq.~(\ref{adelta-f}) into Eq.~(\ref{4.2}) leads to,
	\begin{widetext}   
		\begin{align}\label{4.7}
		(\vec{k_i} \cdot \dot{\vec{\Omega}}_\sigma) \frac{\del f^{0}_i}{\del \om_i} + \frac{\vk_i}{\om_i} \cdot (\vec{k_i} \cdot \frac{\del \vec{\Omega}_{\sigma}}{\del \vec{x}}) \frac{\del f^{0}_i}{\del \om_i} + q_i \left(\vec{E} + \frac{\vk_i}{\om_i} \times \vec{B}\right) \cdot \frac{\del f^{0}_i}{\del \om_i} \frac{\del \om_i}{\del \vec{k_i}} + q_i \left(\vec{E} + \frac{\vk_i}{\om_i} \times \vec{B}\right) \cdot (\vec{\Omega}_{\sigma})\frac{\del \vec{k_i}}{\del \vec{k_i}} \frac{\del f^{0}_i}{\del \om_i}\nn\\
		= -\frac{1}{\tau^i_R}(\vec{k_i} \cdot \vec{\Omega}_{\sigma}) \frac{\del f^{0}_i}{\del \om_i}.
		\end{align}    
		The second term in Eq.~(\ref{4.7}) contributes only to the viscosity and thermal conductivity of the medium but it has a null contribution in the case of electrical conductivity.\\
		The unknown $\vec{\Omega}_\sigma$ from Eq.~(\ref{Omega}) is, 
		\begin{align}\label{aOmega}
		\vec{\Omega}_\sigma = ~\alpha_1 \vec{E} + \alpha_2 \dot{\vec E} +\alpha_3 \vec{B} + \alpha_4 \dot{\vec B} + \alpha_5 (\vec{E} \times \vec{B})+ \alpha_6 (\dot{\vec E} \times \vec{B}) + \alpha_7 (\vec E \times \dot{\vec{B}})~. 
		\end{align}  
		Substitution of Eq.~(\ref{aOmega}) in Eq.~(\ref{4.7}) leads to,  
		\begin{align}\label{4.9}
		&\Big\{\vec{k_i} \cdot\Big(\dot{\alpha_1} \vec{E} + \alpha_1 \dot{\vec{E}} + \dot{\alpha_2 }\dot{\vec E} + \dot{\alpha_5} (\vec{E} \times \vec{B}) + \alpha_5 (\dot{\vec{E}} \times \vec{B}) + \alpha_5 (\vec{E} \times \dot {\vec{B}})  + \dot{\alpha_6} (\dot{\vec E} \times \vec{B}) + \alpha_6 (\dot{\vec E} \times \dot{\vec{B}}) + \dot{\alpha_7} (\vec E \times \dot{\vec{B}}) + \alpha_7 (\dot{\vec E} \times \dot{\vec{B}})\Big)\nn \\
		&+ q_i \vec{E} \cdot \vec{v_i} \cdot  \frac{\del f^{0}_i}{\del \om_i} + q_i \left(\vec{E}+ \frac{\vk_i}{\om_i} \times \vec{B}\right) \cdot (\alpha_1 \vec{E} + \alpha_2 \dot{\vec E} +\alpha_5 (\vec{E} \times \vec{B}) + \alpha_6 (\dot{\vec E} \times \vec{B}) + \alpha_7 (\vec E \times \dot{\vec{B}}))\Big\} \frac{\del f^{0}_i}{\del \om_i}\nn\\ 
		&= -\frac{1}{\tau^i_R}\vec{k_i} \cdot \Big\{(\alpha_1 \vec{E} + \alpha_2 \dot{\vec E} + \alpha_5 (\vec{E} \times \vec{B}) + \alpha_6 (\dot{\vec E} \times \vec{B}) + \alpha_7 (\vec E \times \dot{\vec{B}})\Big\} \frac{\del f^{0}_i}{\del \om_i}.
		\end{align} 
		In this analysis, we consider only the terms with first-order derivatives of the fields and neglect terms with higher-order derivatives. Therefore, the terms with $\dot{\alpha_2 }$, $\dot{\alpha_6}$, $\dot{\alpha_7}$ are ignored in the current analysis. Hence Eq.~(\ref{4.9}) turns to be,
	\end{widetext}
	\begin{widetext}   
		\begin{align}\label{4.10}
		&\Big\{\om_i \vec{v_i} \cdot\Big(\dot{\alpha_1} \vec{E} + \alpha_1 \dot{\vec{E}} + \dot{\alpha_5} (\vec{E} \times \vec{B}) + \alpha_5 (\dot{\vec{E}} \times \vec{B}) + \alpha_5 (\vec{E} \times \dot {\vec{B}}) + \alpha_6 (\dot{\vec E} \times \dot{\vec{B}}) + \alpha_7 (\dot{\vec E} \times \dot{\vec{B}})\Big)\nn \\
		&+ q_i \vec{v_i} \cdot \vec{E} \cdot  \frac{\del f^{0}_i}{\del \om_i} 
		+ q_i \left(\vec{E} \cdot \alpha_1 \vec{E} + \vec{E} \cdot \alpha_2 \dot {\vec{E}}\right) + q_i\alpha_1(\vec{v_i} \times \vec{B} \cdot \vec{E}) + q_i\alpha_2(\vec{v_i} \times \vec{B} \cdot \dot{\vec E}) + q_i\alpha_5(\vec{v_i} \times \vec{B} \cdot (\vec{E} \times \vec{B}))\nn \\
		&+ q_i\alpha_6(\vec{v_i} \times \vec{B} \cdot (\dot{\vec E} \times \vec{B})) 
		+ q_i\alpha_7(\vec{v_i} \times \vec{B} \cdot (\vec E \times \dot{\vec{B}}))\Big\} \frac{\del f^{0}_i}{\del \om_i} \nn\\
		&= -\frac{1}{\tau^i_R}\vec{k_i} \cdot \Big\{\alpha_1 \vec{E} + \alpha_2 \dot{\vec E}+ \alpha_5 (\vec{E} \times \vec{B}) + \alpha_6 (\dot{\vec E} \times \vec{B}) + \alpha_7 (\vec E \times \dot{\vec{B}})\Big\} \frac{\del f^{0}_i}{\del \om_i}.\nn\\
		\Rightarrow~~ &\Big\{\om_i \vec{v_i} \cdot\Big(\dot{\alpha_1} \vec{E} + \alpha_1 \dot{\vec{E}} + \dot{\alpha_5} (\vec{E} \times \vec{B}) + \alpha_5 (\dot{\vec{E}} \times \vec{B}) + \alpha_5 (\vec{E} \times \dot {\vec{B}}) + \alpha_6 (\dot{\vec E} \times \dot{\vec{B}}) + \alpha_7 (\dot{\vec E} \times \dot{\vec{B}})\Big) + q_i \vec{v_i} \cdot \vec{E} \cdot  \frac{\del f^{0}_i}{\del \om_i} \nn \\
		&+ q_i \left(\vec{E} \cdot \alpha_1 \vec{E} + \vec{E} \cdot \alpha_2 \dot {\vec{E}}\right) - q_i\alpha_1(\vec{v_i} \cdot (\vec{E} \times \vec{B})) - q_i\alpha_2(\vec{v_i} \cdot (\dot{\vec{E}} \times \vec{B})) + q_i\alpha_5(- (\vec{v_i} \cdot \vec{B}) (\vec{E} \cdot \vec{B}) + (\vec{v_i} \cdot \vec{E}) (\vec{B} \cdot \vec{B})) \nn\\
		&+ q_i\alpha_6(- (\vec{v_i} \cdot \vec{B}) (\dot{\vec{E}} \cdot \vec{B}) + (\vec{v_i} \cdot \dot{\vec{E}}) (\vec{B} \cdot \vec{B})) 
		+ q_i\alpha_7(- (\vec{v_i} \cdot \dot{\vec{B}}) (\vec{E} \cdot \vec{B}) + (\vec{v_i} \cdot \vec{E}) (\vec{B} \cdot \dot{\vec{B}}))\Big\} \frac{\del f^{0}_i}{\del \om_i} \nn\\
		&= -\frac{\om_i \vec{v_i}}{\tau^i_R} \cdot \Big\{\alpha_1 \vec{E} + \alpha_2 \dot{\vec E}+ \alpha_5 (\vec{E} \times \vec{B}) + \alpha_6 (\dot{\vec E} \times \vec{B}) + \alpha_7 (\vec E \times \dot{\vec{B}})\Big\} \frac{\del f^{0}_i}{\del \om_i}.
		\end{align} 
	\end{widetext}
	Now, we compare the coefficients of same tensor structure on both sides of the above equation one by one. With the comparison of coefficient of $(\vec{v_i} \cdot \vec{E})$ on both sides, we get,\\
	$\om_i \dot{\alpha_1} + q_i + q_i \alpha_5(\vec{B} \cdot \vec{B}) + q_i \alpha_7(\vec{B} \cdot \dot{\vec{B}})$ = $-\frac{\om_i }{\tau^i_R} \alpha_1$,
	
	\begin{align}\label{4.12}
	\Rightarrow~{\dot{\alpha_{1}}} = - \Big\{\frac{1}{\tau^i_R} \alpha_{1} + \frac{q_{i} \alpha_5(\vec{B} \cdot \vec{B})} {\om_i} + \frac{q_{i} \alpha_7(\vec{B} \cdot \dot{\vec{B}})} {\om_i} + \frac{q_{i}}{\om_i}\Big\}. 
	\end{align} 
	The comparison of coefficients of $(\vec{v_i} \cdot \dot{\vec{E}})$, $(\vec{v_i} \cdot \vec{E} \times \vec{B})$, $(\vec{v_i} \cdot \dot{\vec{E}} \times \vec{B})$ and $(\vec{v_i} \cdot \vec{E} \times \dot{\vec{B}})$ gives us the values of $\alpha_2$, $\dot \alpha_5$, $\alpha_6$, and $\alpha_7$ respectively as 
	\begin{align}
	\alpha_2 &= - {\tau^i_R}\Big\{\alpha_{1} + \frac{q_{i} \alpha_6(\vec{B} \cdot \vec{B})} {\om_i}\Big\},\nn \\
	\dot{\alpha_{5}} &= - \frac{1}{\tau^i_R} \alpha_{5} + \frac{q_{i} \alpha_{1}} {\om_i},\nn\\
	\alpha_{6} &= - \tau^i_R(\alpha_{5} - \frac{q_{i} \alpha_{2}} {\om_i}),\nn\\
	\alpha_{7} &= - \tau^i_R\alpha_{5}.\label{4.15}
	\end{align} 
	Substitution of $\alpha_{7}$ from Eq.~(\ref{4.15}) into Eq.~(\ref{4.12}) leads to,
	\begin{align}\label{4.16}
	\dot{\alpha_{1}} = - \Big\{\frac{1}{\tau^i_R} \alpha_{1} + \frac{q_{i}}{\om_i} \left((\vec{B} \cdot \vec{B})  - (\tau^i_R\vec{B} \cdot \dot{\vec{B}})\right)\alpha_5 + \frac{q_{i}}{\om_i}\Big\}. 
	\end{align} 
	These coupled differential equations for $\alpha_{1}$ and $\alpha_{5}$ can be described in the following matrix equation as
	\begin{equation}\label{4.17}
	\frac{d X}{d t}= AX +G.   
	\end{equation}
	where,
	\begin{align}
	X=\begin{pmatrix}
	\alpha_1\\\alpha_2
	\end{pmatrix},  
	~A&=\begin{pmatrix}
	-\frac{1}{\tau_R^i} & -\frac{q_{i} F^2}{\om_i}\\
	\frac{q_{i}}{\om_i} &-\frac{1}{\tau_R^i}\\
	\end{pmatrix},
	G&=\begin{pmatrix}
	-\frac{q_{i}}{\om_i}\\ 0
	\end{pmatrix},\nn
	\end{align}
	with, $F = \sqrt{B(B-\tau_R^i \dot{B})}$.\\ Eq.~(\ref{4.17}) 
	can be solved by diagonalizing the matrix $A$ and using the method of the variation of constants.
	The eigen values corresponding to matrix A are,~\\
	$\lambda _ 1 $ = $-\frac{1}{\tau_R^i} - i\frac{q_{i} F}{\om_i},$
	~~~~~~~$\lambda _ 2 $ = $-\frac{1}{\tau_R^i} + i\frac{q_{i} F}{\om_i}$.\\
	The  eigen vectors corresponding to these eigen values are,
	\begin{align}
	&v_1=\begin{pmatrix}
	-iF\\ 1
	\end{pmatrix},\nonumber
	&v_2=\begin{pmatrix}
	iF\\ 1
	\end{pmatrix}.\nonumber
	\end{align}
	Hence, linear independent solutions corresponding to homogeneous part of differential equation Eq.~(\ref{4.17}) are,
	\begin{align}
	&H_1=\begin{pmatrix}
	-iF e^{\eta_1}\\ e^{\eta_1}
	\end{pmatrix},\nonumber
	&H_2=\begin{pmatrix}
	iF e^{\eta_2}\\ e^{\eta_2}
	\end{pmatrix}.\nonumber
	\end{align}
	Where,
	\begin{align}\label{4.19}
	\eta_j = -\frac{t}{\tau_R^i} +a_j\frac{q_{i} i}{\om_i}\int F dt~,
	\end{align}
	with $a_1=-1$, $a_2=1$.\\
	Therefore, the fundamental matrix for Eq.~(\ref{4.17}) is
	\begin{align}\label{4.20}
	&Y=\begin{pmatrix}
	-iF e^{\eta_1} & iF e^{\eta_2}\\
	e^{\eta_1} & e^{\eta_2}\\
	\end{pmatrix}.
	\end{align}
	We seek a particular solution of equation with given form of Eq.~(\ref{4.17}) is
	\begin{align}\label{4.22}
	Y_p &= YU.
	\end{align}
	Where U is a column matrix of form,
	\begin{align}
	U &=\begin{pmatrix}
	u1\\ u2\\ .\\ .\\ .\\ u_n\\
	\end{pmatrix}.\nonumber
	\end{align}
	We can see that $Y_p$ is a column matrix with same coefficients as that of matrix $X$.
	The differentiation of Eq.~(\ref{4.22}) with respect to time gives us,\\
	$Y_p^{'}$ = $Y^{'}U + YU^{'}$.\\ 
	Here $Y^{'}$ = $AY$. Hence, 
	$Y_p^{'}$ = $AY_p + YU^{'}.$\\
	Comparison of above equation with Eq.~(\ref{4.17}) shows\\ $G$ = $YU^{'}$. \\ 
	The determinant of matrix $Y$ is $-2iFe^{\eta}$. So \\
	\begin{align}\label{4.25}
	u^{'}_1 &= \frac{1}{-2iFe^{\eta}}
	det
	\begin{pmatrix}
	-\frac{q_{i}}{\om_i} & iF e^{\eta_2}\\
	0 & e^{\eta_2}\\
	\end{pmatrix},\nonumber
	\\u^{'}_2 &= \frac{1}{-2iFe^{\eta}} 
	det
	\begin{pmatrix}
	-iF e^{\eta_2} & -\frac{q_{i}}{\om_i}\\
	e^{\eta_2} & 0
	\end{pmatrix}.
	\end{align}
	After integrating both $u^{'}_1$ and $u^{'}_2$ with respect to time, we get the $U$ matrix as\\
	\begin{align}
	U = \frac{q_i}{2i\om_i}
	\begin{pmatrix}
	\int \frac{e^{-\eta_{1}}}{F} dt\\
	-\int \frac{e^{-\eta_{2}}}{F} dt \\
	\end{pmatrix}\nn.
	\end{align}
	\\Substituting the value of $U$ from above equation into  Eq.~(\ref{4.22})
	\begin{align}
	Y_p = 
	\frac{q_i}{2i\om_i}
	\begin{pmatrix}
	-iF e^{\eta_1} & iF e^{\eta_2}\\
	e^{\eta_1} & e^{\eta_2}\\
	\end{pmatrix}
	\begin{pmatrix}
	\int \frac{e^{-\eta_{1}}}{F} dt \\
	-\int \frac{e^{-\eta_{2}}}{F} dt\\
	\end{pmatrix}.\nn
	\end{align}
	With,
	\begin{align}\label{Int}
	I_j &= \int \frac{e^{-\eta_j}}{F}dt,\nn\\
	k_j &= a_j\frac{iqI_j}{2\om}.
	\end{align}
	Hence, we get the particular solution,
	\begin{align}\label{4.28}
	Y_p = 
	\begin{pmatrix}
	\alpha_1\\\alpha_2
	\end{pmatrix} = 
	\begin{pmatrix}
	-ik_1Fe^{\eta_1} + ik_2Fe^{\eta_2}\\
	k_1e^{\eta_1} + k_2e^{\eta_2}\\
	\end{pmatrix}.
	\end{align}
	So,\\
	$\alpha_1 = -ik_1Fe^{\eta_1} + ik_2Fe^{\eta_2}$,
	~~~~~$\alpha_5 = k_1e^{\eta_1} + k_2e^{\eta_2}$.\\
	With, ${\Tilde{\Gamma}_i}=\frac{q_{i}F}{\om_i}$, finally we get, \\
	\begin{align}\label{4.29A}
	\alpha_1  &=-\frac{{\Tilde{\Gamma}}_i}{2}(I_1 e^{\eta_1} +I_2 e^{\eta_2} ),\nn\\
	\alpha_5 &= -\frac{iq_i}{2\om_i} (I_1 e^{\eta_1} - I_2 e^{\eta_2}).
	\end{align}
	Substitution of Eq.~(\ref{4.29A}) in $\alpha_7$ of Eq.~(\ref{4.15}) leads to,
	\begin{align}\label{4.30}
	\alpha_7 &= \frac{i\tau_R^i q_i}{2\om_i} (I_1 e^{\eta_1} - I_2 e^{\eta_2}).
	\end{align}
	With use of Eq.~(\ref{4.15}) and Eq.~(\ref{4.29A}), we obtain,
	\begin{align}\label{4.31}
	\alpha_1 &= -\frac{q_i}{\om_i}  \left(\frac{\frac{1}{\tau_R^i} +\frac{1}{\tau_B}}{(\frac{1}{\tau_R^i} +\frac{1}{\tau_B})^2 + (\frac{\sqrt{1+\frac{\tau_R^i}{\tau_B}}}{\tau_B})^2}\right ),\nn\\
	\alpha_2 &= \frac{(\frac{\tau_R^i{\Tilde{\Gamma}}_i} {2} - \frac{i\tau_R^{i2}{\Gamma}_i^2} {2})I_1 e^{\eta_1} + (\frac{\tau_R{\Tilde{\Gamma}}_i} {2} + \frac{i\tau_R^{i2}{\Gamma}_i^2} {2})I_2 e^{\eta_2} }{1 + \tau_R^{i2}{\Gamma}_i^2}, \nn\\
	\alpha_5 &= -\frac{q_{i}^{2}}{\om_{i}^{2}}  \left(\frac{1}{(\frac{1}{\tau_R^i} +\frac{1}{\tau_B})^2 + (\frac{\sqrt{1+\frac{\tau_R^i}{\tau_B}}}{\tau_B})^2}\right ),\nn\\
	\alpha_6 &= \frac{(\frac{\tau_R^{i2}{\Tilde{\Gamma}}_i^2} {2F} + \frac{iq_i\tau_R^i} {2\om_i})I_1 e^{\eta_1} + (\frac{\tau_R^{i2}{\Tilde{\Gamma}}_i^2} {2F} - \frac{iq_i\tau_R^i} {2\om_i})I_2 e^{\eta_2}}{1 + \tau_R^{i2}{\Gamma}_i^2},\nn\\
	\alpha_7 &= \frac{q_{i}^{2}}{\om_{i}^{2}}  \left(\frac{\tau_R^i}{(\frac{1}{\tau_R^i} +\frac{1}{\tau_B})^2 + (\frac{\sqrt{1+\frac{\tau_R^i}{\tau_B}}}{\tau_B})^2}\right). 
	\end{align}
	In the presence of electromagnetic field the general form of electric current density can be expressed as in Eq.~(\ref{23}), 
	\begin{align}\label{4.23}
	\vec{j} = j_e \hat{{e}} +j_H (\hat{e} \times \hat{{b}})~, 
	\end{align}
	From Eq.~ (\ref{Cur-den}),
	\begin{align}\label{4.24}
	\vec{j} =  \sum_{i} q_i g_i \int \frac{d^3|\vk_i|}{(2\pi)^3} \frac{\vec{k_i}}{\omega_i} f_i~, 
	\end{align}
	with $f_i = f^0_i + \delta f_i$ . \\
	Substituting Eq.~(\ref{adelta-f}) in Eq.~(\ref{4.24}) we get,
	\begin{align}
	{\vec j}&=\sum_{i} q_i g_i \int \frac{d^3|\vk_i|}{(2\pi)^3} \frac{\vec{k_i}}{\vec{\omega_i}}\ (\vec{k_i} \cdot \vec{\Omega}_{\sigma}) \frac{\del f^{0}_i}{\del \om_i}, \nn\\ 
	{\vec j}&= \frac{1}{3}\sum_{i} q_i g_i \int \frac{d^3|\vk_i|}{(2\pi)^3} \frac{\vec{k_i}^{2}}{\vec{\omega_i}} \Big(\alpha_1 \vec{E} + \alpha_2 \dot{\vec E} + \alpha_5 (\vec{E} \times \vec{B}) \nn\\
	&+ \alpha_6 (\dot{\vec E} \times \vec{B}) + \alpha_7 (\vec E \times \dot{\vec{B}})\Big) \frac{\del f^{0}_i}{\del \om_i}.  
	\end{align}
	For the case of a time-evolving magnetic field, we have $j_e=(j_e^{(0)} + j_e^{(1)})$ and $j_H=(j_H^{(0)}+j_H^{(1)}+j_H^{(2)})$, as in Eq.~(\ref{4.23}).
	The components of net current density are of the given form,
	\begin{align}\label{4.26}
	j_e^{(0)} &= \frac{1}{3}\sum_i q_i g_i \int \frac{d^3|\vk_i|}{(2\pi)^3}\frac{\vec{k_i}^{2}}{\vec{\omega_i}} (\alpha_1 \vec{E}) \frac{\del f^{0}_i}{\del \om_i},\nn\\
	j_e^{(1)} &= \frac{1}{3}\sum_i q_i g_i \int \frac{d^3|\vk_i|}{(2\pi)^3}\frac{\vec{k_i}^{2}}{\vec{\omega_i}} (\alpha_2 \dot{\vec E}) \frac{\del f^{0}_i}{\del \om_i},\nn\\
	j_H^{(0)} &= \frac{1}{3}\sum_i q_i g_i \int \frac{d^3|\vk_i|}{(2\pi)^3}\frac{\vec{k_i}^{2}}{\vec{\omega_i}} (\alpha_5 (\vec{E} \times \vec{B})) \frac{\del f^{0}_i}{\del \om_i},\nn\\
	j_H^{(1)} &= \frac{1}{3}\sum_i q_i g_i \int \frac{d^3|\vk_i|}{(2\pi)^3}\frac{\vec{k_i}^{2}}{\vec{\omega_i}} (\alpha_6 (\dot{\vec E} \times \vec{B})) \frac{\del f^{0}_i}{\del \om_i},\nn\\
	j_H^{(2)} &= \frac{1}{3}\sum_i q_i g_i \int \frac{d^3|\vk_i|}{(2\pi)^3}\frac{\vec{k_i}^{2}}{\vec{\omega_i}} (\alpha_7 (\vec E \times \dot{\vec{B}})) \frac{\del f^{0}_i}{\del \om_i}.
	\end{align}
	With use of Eq.~(\ref{4.31}), we get
	\begin{align}\label{App-Elec-Cur}
	&j_e^{(0)} = \frac{E}{3T} \sum_i g_i (q_{i})^2 \int \frac{d^3|\vk_i|}{(2\pi)^3}
	\frac{ \vk^2_i}{\om_i^2} f^0_i(1\mp f^0_i) ~ \tau_R^i \nn\\
	&~~~~~~~~~~~~~~~~~~~~~~~~~~~~ \times \frac{1}{1+\chi_i+\chi_i^2}, \nn\\
	\nn &j_e^{(1)} = -\frac{\dot{E}}{3T} \sum_i g_i (q_{i})^2 \int \frac{d^3|\vk_i|}{(2\pi)^3}
	\frac{ \vk^2_i}{\om_i^2} f^0_i(1\mp f^0_i)~ {\tau_R^i}^2 \nn\\
	\nn &~~~~~~~~~~~~~~~~~~~~~~~~~ \times \frac{1+\chi_i-\chi_i^2}{(1+\chi_i)(1+\chi_i^2)(1+\chi_i+\chi_i^2)}, \\
	\nn &j_H^{(0)} = \frac{E B}{3T} \sum_i g_i (q_{i})^3 \int \frac{d^3|\vk_i|}{(2\pi)^3}
	\frac{ \vk^2_i}{\om_i^3} f^0_i(1\mp f^0_i)~ {\tau_R^i}^2\\
	\nn &~~~~~~~~~~~~~~~~~~~~~~~~~~~~ \times \frac{1}{(1+\chi_i)(1+\chi_i+\chi_i^2)}, \\ 
	\nn &j_H^{(1)} = -\frac{\dot{E}B}{3T} \sum_i g_i (q_{i})^3 \int \frac{d^3|\vk_i|}{(2\pi)^3}
	\frac{ \vk^2_i}{\om_i^3} f^0_i(1\mp f^0_i) ~ {\tau_R^i}^3\\
	\nn &~~~~~~~~~~~~~~~~~~~~~~~~~ \times \frac{\chi_i}{(1+\chi_i)(1+\chi_i^2)(1+\chi_i+\chi_i^2)}, \\
	&j_H^{(2)} = -\frac{E\dot{B}}{3T} \sum_i g_i (q_{i})^3 \int \frac{d^3|\vk_i|}{(2\pi)^3}
	\frac{ \vk^2_i}{\om_i^3} f^0_i(1\mp f^0_i) ~ {\tau_R^i}^3 \nn\\
	&~~~~~~~~~~~~~~~~~~~~~~~~~~~~ \times \frac{1}{(1+\chi_i)(1+\chi_i+\chi_i^2)}.
	\end{align}
	From Eq.~(\ref{C-EC}) and Eq.~(\ref{App-Elec-Cur}), components of electrical conductivity can be found as
	\begin{align}
	\sigma^{(0)}_e &= \frac{1}{3T} \sum_i g_i (q_{i})^2 \int \frac{d^3|\vk_i|}{(2\pi)^3}
	\frac{ \vk^2_i}{\om_i^2} f^0_i(1\mp f^0_i) ~ \tau_R^i \nn\\
	\nn &~~~~~~~~~~~~~~~~~~~~~~~~~~~~ \times \frac{1}{1+\chi_i+\chi_i^2}, \\
	\sigma^{(1)}_e &= \frac{1}{3T} \sum_i g_i (q_{i})^2 \int \frac{d^3|\vk_i|}{(2\pi)^3}
	\frac{ \vk^2_i}{\om_i^2} f^0_i(1\mp f^0_i) ~\frac{{\tau_R^i}^2}{\tau_E} \nn\\
	\nn &~~~~~~~~~~~~~~~~~~~~~~~ \times \frac{1+\chi_i-\chi_i^2}{(1+\chi_i)(1+\chi_i^2)(1+\chi_i+\chi_i^2)}, \\
	\sigma^{(0)}_H &= \frac{1}{3T} \sum_{i} g_i (q_{i})^2 \int \frac{d^3|\vk_i|}{(2\pi)^3}
	\frac{ \vk^2_i}{\om_i^2} f^0_i(1\mp f^0_i) ~{{\tau_R^i}^2 \Gamma_i} \nn\\
	\nn &~~~~~~~~~~~~~~~~~~~~~~~~~~~~ \times \frac{1}{(1+\chi_i)(1+\chi_i+\chi_i^2)}, \\
	\sigma^{(1)}_H &= \frac{1}{3T} \sum_{i} g_i (q_{i})^2 \int \frac{d^3|\vk_i|}{(2\pi)^3}
	\frac{ \vk^2_i}{\om_i^2} f^0_i(1\mp f^0_i) \frac{{\tau_R^i}^3 \Gamma_i}{\tau_E} \nn\\
	\nn &~~~~~~~~~~~~~~~~~~~~~~~~~ \times \frac{\chi_i}{(1+\chi_i)(1+\chi_i^2)(1+\chi_i+\chi_i^2)},\\ 
	\sigma^{(2)}_H &= \frac{1}{3T} \sum_{i} g_i (q_{i})^2 \int \frac{d^3|\vk_i|}{(2\pi)^3}
	\frac{ \vk^2_i}{\om_i^2} f^0_i(1\mp f^0_i) \frac{{\tau_R^i}^3 \Gamma_i}{\tau_B} \nn\\
	&~~~~~~~~~~~~~~~~~~~~~~~~~ \times \frac{1}{(1+\chi_i)(1+\chi_i+\chi_i^2)}~,
	\label{ App-ele-cond}
	\end{align}
	with $\Gamma_i = \frac{q_i B}{\om_i}$.
	%
	\setcounter{equation}{0}
	\subsection{Appendix: Thermal Conductivity}
	\label{appendix2}
	The general form of ${\vec{\Omega}_\kappa}$ from Eq.~(\ref{1.7}) is
	\begin{align}\label{5.9}
	\vec{\Omega}_\kappa =& \alpha_1\vec{B}+ \alpha_2\vec{\nabla}T+ \alpha_3(\vec{\nabla}T \times \vec{B})+\alpha_4 \dot{\vec{B}} +\alpha_5(\vec{\nabla}T \times \dot{\vec{B}})\nonumber \\
	=& \alpha_1\vec{B}+ \alpha_2\vec{\nabla}T +\alpha_4 \dot{\vec{B}} + (\alpha_3 \lvert \vec{B} \rvert + \alpha_5
	\lvert \dot{\vec{B}} \rvert) (\vec{\nabla}T\times \hat{b}).
	\end{align} 
	The unknown coefficients $\alpha_{i}$ ($i=(1, 2,.., 5)$) can be obtained by substituting Eq.~(\ref{5.9}) in the Boltzmann equation. 
	The Boltzmann equation under the relaxation time approximation 
	\begin{align}\label{5.9A}
	\frac{\del f_i}{\del t} + \frac{\vk_i}{\om_i} \cdot \frac{\del f_i}{\del \vec{x}} + q_i \left(\frac{\vk_i}{\om_i} \times \vec{B}\right) \cdot \frac{\del f_i}{\del \vec{k_i}}
	= -\frac{\delta f_i}{\tau^i_R}~.
	\end{align}
	Employing Eq.~(\ref{5.9}) in Eq.~(\ref{5.9A}),
	The first term in left hand side (lhs) of the equation becomes,
	\begin{align}\label{5.10}
	& \om_i {\vec{v}_i}.\Big\{\dot{\alpha_1 }{\vec {B}}+\alpha_1 \dot{{\vec {B}}}+\dot{\alpha_2} {\vec {\nabla}T}+\alpha_3 ({\vec {\nabla}T}\times \dot{{\vec {B}}}) +\dot{\alpha_3 }({\vec {\nabla}T} \times {\vec {B}})\nonumber\\
	&+\dot{\alpha_4} \dot{\vec{B}}+\alpha_4 \ddot{\vec{B}}+\dot{\alpha_5 }({\vec {\nabla}T} \times \dot{{\vec {B}}}) +\alpha_5 (\vec{\nabla}T \times\ddot{\vec{B}})\Big\}.
	\end{align}
	The second term in lhs of the Boltzmann equation leads to, \\
	\begin{align}\label{5.11}
	\frac{\partial f_i}{\partial x_i} &= \frac{\partial}{\partial x_i}(f^{0}_i +\delta f_i)\nonumber\\ &=\frac{\partial}{\partial x_i}(f^0_i) + \frac{\partial f_i}{\partial x_i}(\delta f_i) \nn\\
	&=\frac{\partial}{\partial x_i}\Big\{\frac{1}{1\pm\exp{\big(\beta( \om_i - {\rm b}_i \mu_i)\big)}}\Big\} + 0\nonumber\\
	&= -(\om_i - {\rm b}_ih)\frac{\vec{\nabla}T}{T}\frac{\partial f^0_i}{\partial \om_i}.
	\end{align}
	Here, in the above equation, we have ignored the contributions of shear and bulk viscosity components.
	The third term in lhs leads to,
	\begin{align}\label{5.12}
	\frac{\partial f_i}{\partial k_i} &= \frac{\partial f^0_i}{\partial k_i} + \frac{\partial \delta f_i}{\partial k_i}\nonumber\\ 
	&= \vec{v_i}\frac{\partial f^0_i}{\partial \om_i} + 
	{\vec {\Omega}_\kappa}  \frac{\partial f^0_i}{\partial \om_i}.
	\end{align}
	The identity $({\vec {v_i}} \times {\vec {B}}) \cdot \vec{v_i}\frac{\partial f^0_i}{\partial \om_i} = 0$. Thus we are left with only ${\vec {\Omega}_\kappa}  \frac{\partial f^0_i}{\partial \om_i}$ term of the above equation. Hence,
	\begin{align}\label{5.13}
	\frac{\partial f_i}{\partial k_i} &= -\alpha_2 q_{i}\vec{v_i}\cdot(\vec{\nabla}T \times \vec{B}) +\alpha_3 q_{i}\vec{v_i}\cdot\vec{\nabla}T(\vec{B}\cdot\vec{B})-\alpha_3 q_{i}\nonumber\\
	\vec{v_i}\cdot\vec{B}
	&(\vec{B}\cdot\vec{\nabla}T)
	+\alpha_5q_{i}\vec{v_i}\cdot\vec{\nabla}T(\dot{\vec{B}}\cdot\vec{B}) -\alpha_5 q_{i}\vec{v_i}\cdot\dot{\vec{B}}(\vec{B}\cdot\vec{\nabla}T).  \end{align}
	Finally, after the substitution of above results in both sides of Eq.~(\ref{5.9A}), we get
	\begin{align}\label{5.14}
	&\om_i {\vec{v_i}}.\Big\{\dot{\alpha_1 }{\vec {B}}+\alpha_1 \dot{{\vec {B}}}+\dot{\alpha_2} {\vec {\nabla}T}+\alpha_3 ({\vec {\nabla}T}\times \dot{{\vec {B}}}) +\dot{\alpha_3 }({\vec {\nabla}T} \times {\vec {B}})
	\nonumber\\
	&+\dot{\alpha_4} \dot{\vec{B}}+\alpha_4\ddot{\vec{B}}+\dot{\alpha_5 }({\vec {\nabla}T} \times \dot{{\vec{B}}}) +\alpha_5 (\vec{\nabla}T\times\ddot{\vec{B}})\Big\}\nonumber\\
	&-(\om_i -{\rm b}_i h)\vec{v_i}\cdot\frac{\vec{\nabla}T}{T}
	-\alpha_2q_{i}\vec{v_i}\cdot(\vec{\nabla}T \times \vec {B}) +\alpha_3q_{i}\vec{v_i}\cdot\vec{\nabla}T\nn\\
	&(\vec{B}\cdot\vec{B})
	-\alpha_3 q_{i}\vec{v_i}\cdot\vec{B}(\vec{B}\cdot\vec{\nabla}T)
	+\alpha_5q_{i}\vec{v_i}\cdot\vec{\nabla}T(\dot{\vec{B}}\cdot\vec{B}) -\alpha_5 q_{i}\nonumber\\
	&\vec{v_i}\cdot\dot{\vec{B}}
	(\vec{B}\cdot\vec{\nabla}T)
	=-\frac{\om_i}{\tau_R^i}\Big\{\alpha_1 {\vec {v_i}}\cdot{\vec B}
	+\alpha_2 {\vec {v_i}}\cdot{\vec {\nabla}T} \nn\\
	&+\alpha_3 {\vec {v_i}}\cdot({\vec {\nabla}T}\times {\vec {B}}) 
	+\alpha_4 {\vec{v_i}}\cdot\dot{\vec{B}}
	+\alpha_5 {\vec {v_i}}\cdot({\vec {\nabla}T}\times \dot{{\vec {B}}})\Big\}.
	\end{align}
	In current analysis, we consider only the terms with first-order derivatives of the fields and neglect higher-order derivative terms.
	Let us now compare the coefficients of same tensor structure on both sides of above equation one by one. The comparison of the coefficients of ${\vec {v}}\cdot{\vec {B}}$ on both sides  leads to,
	\begin{align}\label{dl1}
	\dot{\alpha_1} &= -\frac{1}{\tau_R^i}\alpha_1+\frac{q_{i}(\vec{B}\cdot\vec{\nabla}T)}{\om_i}\alpha_3. 
	\end{align}
	Similarly the comparison of coefficients of ${\vec {v_i}\cdot \dot{\vec{B}}}$, ${\vec {v_i}}\cdot{\vec {\nabla}T}$, ${\vec {v_i}}\cdot({\vec {\nabla}T \times \dot{\vec{B}}})$, and ${\vec {v_i}}\cdot({\vec {\nabla}T\times \vec{B}})$ gives us $\alpha_4$, $\dot{\alpha_2}$, $\alpha_5$, $\dot{\alpha_3}$ respectively as
	\begin{align}\label{dot4}
	{\alpha_4} &=-{\tau_R^i}\Big\{\alpha_1+\frac{\tau_R^i q_{i} (\vec{B}\cdot\vec{\nabla}T)}{\om_i} \alpha_5 \Big\},\nn\\ 
	\dot{\alpha_2} &=  -\bigg\{\frac{1}{\tau_R^i}\alpha_2 +(\frac{q_{i}(\vec{B}\cdot\vec{B}-\tau_R^i\vec{B}\cdot\dot{\vec{B}})}{\om_i})\alpha_3 -\frac{\om_i - {\rm b}_ih}{T\om_i}\bigg\},\nn \\
	\dot{\alpha_3} &= -\frac{1}{\tau_R^i}\alpha_3 +\frac{q_{i}}{\om_i}\alpha_2,\nn \\
	\alpha_5 &= -\tau_R^i \alpha_3.  
	\end{align}
	Here, $\dot{\alpha_1}$ from Eq.~(\ref{dl1}) and $\dot{\alpha_2}$, $\dot{\alpha_3}$ Eq.~(\ref{dot4}) can be expressed in terms of matrix equation as
	\begin{equation}\label{5.16}
	\frac{d X}{d t}= AX +G,   
	\end{equation}
	where the matrices take the following forms,
	\begin{align}
	X=\begin{pmatrix}
	\alpha_1\\\alpha_2\\\alpha_3
	\end{pmatrix},  
	A=\begin{pmatrix}
	-\frac{1}{\tau_R^i} &0 &\frac{q_{i}}{\om_i}(\vec{B}\cdot\vec{\nabla}T)\\
	0 &-\frac{1}{\tau_R^i} & -\frac{q_{i} F^2}{\om_i}\\
	0 &\frac{q_{i}}{\om_i} &  -\frac{1}{\tau_R^i}, 
	\end{pmatrix},
	G =\begin{pmatrix}
	0\\ \frac{\om_i - {\rm b}_ih}{T\om_i}\\ 0
	\end{pmatrix},\nonumber
	\end{align}
	with $F = \sqrt{B(B-\tau_R^i \dot{B})}$. Eq.~(\ref{5.16}) can be solved by diagonalizing the matrix $A$ and using the method of the variation of constants. The eigen values corresponding to matrix $A$ are,\\
	$\lambda _ j $ = $-\frac{1}{\tau_R^i} + a_j i\frac{q_{i} F}{\om_i}$,~~
	with $a_1=0$, $a_2=-1$, $a_3=1$.\\
	The eigen vectors corresponding to these eigen values are,
	\begin{align}
	v_1=\begin{pmatrix}
	1\\ 0\\ 0
	\end{pmatrix},
	v_2=\begin{pmatrix}
	\frac{-(\vec{B}\cdot\vec{\nabla}T)}{iF}\\-iF\\ 1
	\end{pmatrix},      
	v_3=\begin{pmatrix}
	\frac{(\vec{B}\cdot\vec{\nabla}T)}{iF}\\iF\\ 1
	\end{pmatrix}.\nonumber
	\end{align}
	Hence, the linear independent solutions corresponding to homogeneous part of differential  Eq.~ (\ref{5.16})
	\begin{align}
	y_1=\begin{pmatrix}
	e^{\eta_1}\\ 0 \\ 0
	\end{pmatrix},
	y_2=\begin{pmatrix}
	\zeta e^{\eta_2} \\-iF e^{\eta_2}\\ e^{\eta_2}
	\end{pmatrix},
	y_3=\begin{pmatrix}
	-\zeta e^{\eta_3} \\iF e^{\eta_3}\\ e^{\eta_3}
	\end{pmatrix}.\nonumber
	\end{align}
	Where,
	\begin{align}\label{5.19}
	\zeta = \frac{i(\vec{B}\cdot\vec{\nabla}T)}{F},
	~\eta_j = -\frac{t}{\tau_R^i} +a_j\frac{q_{i} i}{\om}\int F dt.
	\end{align}
	Therefore, the fundamental matrix for Eq.~(\ref{5.16}) is 
	\begin{align}\label{5.20}
	&Y=\begin{pmatrix}
	e^{\eta_1}& \zeta e^{\eta_2} & -\zeta e^{\eta_3}\\
	0 & -iF e^{\eta_2} & iF e^{\eta_3}\\
	0 & e^{\eta_2} & e^{\eta_3}\\
	\end{pmatrix},
	\end{align}
	We seek a particular solution of equation with given form of Eq.~(\ref{5.16})
	is
	\begin{align}\label{5.22}
	Y_p &= YU
	\end{align}
	where $U$ is a column matrix of unknowns. 
	\begin{align}\label{5.23}
	U=\begin{pmatrix}
	u1\\ u2\\ .\\ .\\ .\\ u_n\\
	\end{pmatrix}.
	\end{align}
	From Eq.~(\ref{5.22}) and Eq.~(\ref{5.16}) we can see that $Y_p$ is a column matrix with same coefficients as that of matrix X. Further, the differentiation of Eq.~(\ref{5.22}) with respect to time gives us
	$Y_p^{'}$ = $Y^{'}U + YU^{'}$, where $Y^{'} = AY$. 
	Hence, 
	$Y_p^{'}$ = $AY_p + YU^{'}$.
	Comparison of above equation with Eq.~(\ref{5.16}) tells us 
	\begin{align}\label{5.24}
	G = YU^{'}.   
	\end{align}
	The determinant of matrix $Y$ is $-2iFe^{\eta}$. Then, 
	\begin{align}\label{5.25}
	u^{'}_1 &= \frac{1}{-2iFe^{\eta}}
	det
	\begin{pmatrix}
	0& \zeta e^{\eta_2} & -\zeta e^{\eta_3}\\
	\frac{\om_i - {\rm b}_ih}{T\om} & -iF e^{\eta_2} & iF e^{\eta_3}\\
	0 & e^{\eta_2} & e^{\eta_3}\\
	\end{pmatrix}, \nonumber\\
	&= \frac{\zeta e^{-\eta_1}}{iF} (\frac{\om_i - {\rm b}_ih}{T\om_i}). \nonumber
	\\u^{'}_2 &= \frac{1}{-2iFe^{\eta}} 
	det
	\begin{pmatrix}
	e^{\eta_1}& 0 & -\zeta e^{\eta_3}\\
	0 & \frac{\om_i - {\rm b}_ih}{T\om_i} & iF e^{\eta_3}\\
	0 & 0 & e^{\eta_3}\\
	\end{pmatrix},\nonumber\\
	&= \frac{-e^{-\eta_2}}{iF} (\frac{\om_i - {\rm b}_ih}{2T\om_i}).\nonumber
	\\u^{'}_3 &= \frac{1}{-2iFe^{\eta}}
	det
	\begin{pmatrix}
	e^{\eta_1}& \zeta e^{\eta_2} & 0\\
	0 & -iF e^{\eta_2} & \frac{\om_i - {\rm b}_ih}{T\om_i}\\
	0 & e^{\eta_2} & 0\\
	\end{pmatrix},\nonumber\\
	&= \frac{e^{-\eta_3}}{iF} (\frac{\om_i -{\rm b}_ih}{2T\om_i}). 
	\end{align}
	After integrating $u^{'}_1$, $u^{'}_2$ and $u^{'}_3$ with respect to time , we get the matrix $U$ as\\
	\begin{align}\label{5.26}
	U &=
	\begin{pmatrix}
	-\zeta i \frac{\om_i - {\rm b}_ih}{T\om_i} \xi_1\\
	i \frac{\om_i - {\rm b}_ih}{2T\om_i} \xi_2\\
	-i \frac{\om_i - {\rm b}_ih}{2T\om_i} \xi_3\\
	\end{pmatrix}.
	\end{align}
	Where $\xi_j = \int \frac{e^{-\eta_{j}}}{F} dt$. \\
	\\After substitution the above value of $U$ in Eq.~(\ref{5.22}) results in,
	
	\begin{align}
	Y_p &= 
	\begin{pmatrix}
	e^{\eta_1}& \zeta e^{\eta_2} & -\zeta e^{\eta_3}\\
	0 & -iF e^{\eta_2} & iF e^{\eta_3}\\
	0 & e^{\eta_2} & e^{\eta_3}\\
	\end{pmatrix}
	\begin{pmatrix}
	-\zeta i \frac{\om_i - {\rm b}_ih}{T\om_i} \xi_1\\
	i \frac{\om_i - {\rm b}_ih}{2T\om_i} \xi_2\\
	-i \frac{\om_i - {\rm b}_ih}{2T\om_i} \xi_3\\
	\end{pmatrix}, \nonumber\\
	\begin{pmatrix}
	\alpha_1\\\alpha_2\\\alpha_3
	\end{pmatrix}
	&= 
	\begin{pmatrix}
	\zeta e^{\eta_1}& \zeta e^{\eta_2} & -\zeta e^{\eta_3}\\
	0 & -iF e^{\eta_2} & iF e^{\eta_3}\\
	0 & e^{\eta_2} & e^{\eta_3}\\
	\end{pmatrix} 
	\begin{pmatrix}
	c_1\\
	c_2\\
	c_3\\
	\end{pmatrix}.\nonumber
	\end{align}
	Hence,\\
	\begin{align}\label{talpha3}
	\alpha_1 &= c_1\zeta e^{\eta_1} + c_2 \zeta e^{\eta_2} - c_3 \zeta e^{\eta_3}, \nonumber\\
	\alpha_2 &= -c_2iFe^{\eta_2} + c_3iFe^{\eta_3},\nonumber\\
	\alpha_3 &= c_2e^{\eta_2} + c_3e^{\eta_3}.   
	\end{align}
	The functions $c_1(t)$, $c_2(t)$ and $c_3(t)$ can be defined as $c_1 =-i\frac{(\om_i - {\rm b}_ih)}{T\om_i}  \xi_1$,  $c_2 =i\frac{(\om_i - {\rm b}_ih)}{2T\om_i} \xi_2$ and $c_3 =-i\frac{(\om_i - {\rm b}_ih)}{2T\om_i} \xi_3$ 
	Here, $ B =  B_0 e^{-\frac{t}{\tau_B}}$, and $\tau_B$ is used as a parameter. For time varying field we get\\
	$ F = B\sqrt{1+\frac{\tau_R^i}{\tau_B}}$. \\
	Hence, it leads to the given form of, 
	\begin{align}\label{5.29}
	\eta_j &= -\frac{t}{\tau_R^i} +a_ji\frac{\sqrt{1+\frac{\tau_R^i}{\tau_B}}}{\tau_B}~t,\nn \\ 
	\xi_j &= \frac{1}{\sqrt{1+\frac{\tau_R^i}{\tau_B}} B_0} \frac{e^{\Big(\frac{1}{\tau_R^i} +\frac{1}{\tau_B}-a_ji\frac{\sqrt{1+\frac{\tau_R^i}{\tau_B}}}{\tau_B} \Big)t}}{\Big( \frac{1}{\tau_R^i} +\frac{1}{\tau_B}-a_ji\frac{\sqrt{1+\frac{\tau_R^i}{\tau_B}}}{\tau_B} \Big)}.
	\end{align}
	The microscopic definition of heat flow can take from Eq.~(\ref{1.5a}),
	\begin{align}
	{\vec {I}}_i &= \int \frac{d^3|\vk_i|}{(2\pi)^3} \frac{\vec {k}_i}{\om_i} (\om_i -{\rm b}_i h)\delta f_i.\nn\\
	&= \frac{1}{3}\sum_i \int \frac{d^3|\vk_i|}{(2\pi)^3\om_{i}} {\vec {k}}_i^2 (\om_i -{\rm b}_ih)\Big\{\alpha_1\vec{B}+ \alpha_2\vec{\nabla}T\nn\\
	&+\alpha_3(\vec{\nabla}T\times \vec{B})
	+\alpha_4 \dot{\vec{B}} +\alpha_5(\vec{\nabla}T\times \dot{\vec{B}})\Big\}(\frac{\partial f_i^0}{\partial \om_i}).\label{5.32}
	\end{align}\\
	In the presence of time-varying magnetic field heat current in the fluid rest frame can be expressed as~\cite{PhysRevD.106.034008}
	\begin{align}\label{5.I}
	\vec{I} &= \kappa_0 \vec{\nabla}T +  (\kappa_1+\kappa_2) (\vec{\nabla}T \times \hat{b})\nonumber \\ 
	&= \kappa_0 \vec{\nabla}T +  \kappa_H (\vec{\nabla}T \times \hat{b})~.   
	\end{align}
	Compare the coefficients of Eq.~(\ref{5.I}) and Eq.~(\ref{5.32})
	\begin{align}
	\kappa_0 &=  \frac{1}{3} \sum_{\rm baryons} g_i \int \frac{d^3|\vk_i|}{(2\pi)^3\om_{i}} k^2_i(\om_i - {\rm b}_ih) \alpha_2(\frac{\partial f_i^0}{\partial \om_i}),\nn\\ 
	\kappa_1 &=  \frac{1}{3} \sum_{\rm baryons} g_i \int \frac{d^3|\vk_i|}{(2\pi)^3\om_{i}} k^2_i(\om_i - {\rm b}_ih) \alpha_3(\frac{\partial f_i^0}{\partial \om_i}),\nn \\
	\kappa_2 &=  \frac{1}{3} \sum_{\rm baryons} g_i \int \frac{d^3|\vk_i|}{(2\pi)^3\om_{i}} k^2_i(\om_i - {\rm b}_ih) (-\tau_R^i \alpha_3) (\frac{\partial f_i^0}{\partial \om_i})\label{5.33C}. 
	\end{align}
	where,
	\begin{align}
	\alpha_2 = \frac{-(\om_i - {\rm b}_i h)}{T\om_i} \frac{\frac{1}{\tau_R^i} + \frac{1}{\tau_B} }{\Big(\frac{1}{\tau_R^i} + \frac{1}{\tau_B}\Big)^2 + \Big(\frac{\sqrt{1+\frac{\tau_R^i}{\tau_B}}}{\tau_B}\Big)^2},  \nn \\ 
	\alpha_3 = \frac{(\om_i - {\rm b}_ih)}{T\om_i B\tau_B } \frac{1}{\Big(\frac{1}{\tau_R^i} + \frac{1}{\tau_B}\Big)^2 + \Big(\frac{\sqrt{1+\frac{\tau_R^i}{\tau_B}}}{\tau_B}\Big)^2}.  \label{alphathree}
	\end{align}
	Substitute value of $\alpha_2$ and $\alpha_3$ from Eq.~(\ref{alphathree}) into Eq.~(\ref{5.33C}) 
	\begin{widetext}
		\begin{align}\label{}
		&\kappa_0 =  \frac{1}{3T^2} \sum_i g_i \int \frac{d^3|\vk_i|}{(2\pi)^3}\frac{\vec{k}^2_i}{\om_i^2}(\om_i - {\rm b}_i h)^2 \tau_R^i ~\frac{1}{(1+\chi_i + \chi_i^2)}~ f^0_i(1\mp f^0_i),\nn\\
		&{\kappa}_1 = \frac{1}{3T^2} \sum_i g_i \int \frac{d^3|\vk_i|}{(2\pi)^3}\frac{ \vec{k}^2_i}{\om_i^2}(\om_i - {\rm b}_i h)^2 \tau_R^i ~\frac{\chi_i}{(1+\chi_i)(1+\chi_i + \chi_i^2)} ~f^0_i(1\mp f^0_i), \nn\\
		&{\kappa}_2 = \frac{1}{3T^2} \sum_i g_i \int \frac{d^3|\vk_i|}{(2\pi)^3}\frac{\vec {k}^2_i}{\om_i^2}(\om_i - {\rm b}_i h)^2 \tau_R^i ~\frac{\chi_i^2}{(1+\chi_i)(1+\chi_i + \chi_i^2)} ~f^0_i(1\mp f^0_i), \nn\\
		{\rm or,}\nn\\
		&{\kappa}_H = \kappa_1 + \kappa_2 = \frac{1}{3T^2} \sum_i g_i \int \frac{d^3|\vk_i|}{(2\pi)^3}\frac{\vec {k}^2_i}{\om_i^2}(\om_i - {\rm b}_i h)^2 \tau_R^i ~\frac{\chi_i}{(1+\chi_i + \chi_i^2)} ~f^0_i(1\mp f^0_i).
		\end{align}
		where $\chi_i = \frac{\tau_R^i}{\tau_B}$.
	\end{widetext}
	
	\bibliographystyle{apsrev4-2}
	\bibliography{reference}
	
\end{document}